\documentclass[acmsmall,screen,nonacm]{acmart}
\AtBeginDocument{%
  \providecommand\BibTeX{{%
    \normalfont B\kern-0.5em{\scshape i\kern-0.25em b}\kern-0.8em\TeX}}}

\setcopyright{acmcopyright}
\copyrightyear{2026}
\acmYear{2026}
\usepackage{multirow}
\usepackage{colortbl}
\usepackage{rotating}
\usepackage{framed}
\usepackage{listings}
\usepackage{xcolor}
\usepackage{booktabs}
\usepackage{tabularx}
\usepackage{array}
\usepackage{seqsplit}
\usepackage{threeparttable}
\usepackage{tcolorbox}
\usepackage{makecell}
\usepackage{tikz}
\usepackage{pgfplots}
\pgfplotsset{compat=1.18}
\newcommand{\codeid}[1]{\texttt{\seqsplit{#1}}}
\newtcolorbox{keyfindingbox}[1]{
  colback=white,
  colframe=black,
  coltitle=white,
  colbacktitle=black,
  title=\textbf{#1},
  fonttitle=\bfseries,
  arc=2pt,
  boxrule=0.8pt,
  left=6pt,
  right=6pt,
  top=5pt,
  bottom=5pt,
  titlerule=0pt,
  toptitle=3pt,
  bottomtitle=3pt
}

\begin{document}

\title{LLM-Enhanced Commit Message Generation via Issue Information: An Exploratory Study}


\author{Zongen Ren}
\orcid{0009-0002-9924-543X}
\affiliation{%
  \institution{School of Computer Science, Wuhan University}
  \city{Wuhan}
  \country{China}}
\email{zongnren@whu.edu.cn}

\author{Wei Shi}
\orcid{0009-0002-8371-9266}
\affiliation{%
  \institution{School of Computer Science, Wuhan University}
  \city{Wuhan}
  \country{China}}
\email{shiwei.acc@gmail.com}

\author{Bo Xiong}
\orcid{0009-0007-4559-7815}
\affiliation{%
  \institution{School of Computer Science, Wuhan University}
  \city{Wuhan}
  \country{China}}
\email{yueshaomoon_@whu.edu.cn}

\author{Chong Wang}
\orcid{0000-0003-4576-5392}
\affiliation{%
  \institution{School of Computer Science, Wuhan University}
  \city{Wuhan}
  \country{China}}
\email{cwang@whu.edu.cn}

\author{Peng Liang}
\orcid{0000-0002-2056-5346}
\affiliation{%
  \institution{School of Computer Science, Wuhan University}
  \city{Wuhan}
  \country{China}}
\email{liangp@whu.edu.cn}
\renewcommand{\shortauthors}{Ren et al.}


\begin{abstract}
In software development and maintenance, commit messages help developers understand code changes, support collaboration, and improve long-term maintenance. To improve the quality of commit messages generated automatically, previous studies on commit message generation (CMG) increasingly focused on supplying Large Language Models (LLMs) with richer contextual information. However, the use of issue information alone as the external context for LLM-based CMG has not been systematically studied. To this end, we aim to enhance the performance of LLM-based CMG by integrating issue information with code diffs, thereby generating commit messages that are more precise and informative. We propose an \textbf{IS}sue-\textbf{A}ugmented framework for \textbf{C}ommit message generation (ISAC) by combining code diffs with issue information as LLM input. To support the evaluation, we construct ApacheCM-Issue, a commit-issue aligned dataset built upon ApacheCM by linking commits with issues from GitHub and Apache Jira. To evaluate the effects of incorporating issue information, adding similar historical commits, and replacing raw issue information with structured summaries, we compare four input configurations across popular LLMs: code diffs only, code diffs with issue information, code diffs with issue information and similar historical commits from the same project, and code diffs with structured issue summaries instead of raw issue text. In addition, we compare ISAC with four CMG baselines and conduct a human evaluation of clarity, completeness, and correctness on 50 stratified samples. The results show that incorporating issue information consistently improves LLM-based CMG across all evaluated model configurations and metrics. On average, issue augmentation improves BLEU, ROUGE-L, METEOR, CIDEr, and SBERT-Cos by 23.60\%, 20.48\%, 25.21\%, 30.70\%, and 8.70\%, respectively. Incorporating a similar historical commit further improves these metrics by 29.49\%, 14.84\%, 13.67\%, 41.08\%, and 2.66\%. Compared with using the full issue information, replacing it with a structured issue summary decreases the five metrics by 9.90\%, 6.99\%, 9.86\%, 10.65\%, and 3.46\%, respectively. ISAC also outperforms the four reproduced baseline methods across all five automatic metrics on the experimental dataset. The human evaluation shows that original issue information improves the clarity of commit messages across all model configurations and generally enhances completeness while maintaining correctness. Human evaluation indicates that structured issue summaries can improve the perceived completeness of commit messages, but replacing the original issue information may sacrifice contextual details and yield worse results on automatic metrics.
\end{abstract}





\maketitle

\section{Introduction}
A commit message is a textual record in a version control system that describes what a code diff changes, why the change was made, and what it affects. A high-quality commit message summarizes the code diff while revealing the task background, the underlying problem, and the intended effect. Such messages are essential for supporting developers in code review, defect tracking, version rollback, and long-term maintenance activities. In continuously evolving projects, commit messages and code diffs together serve as crucial evidence to understand the historical development of a software system. In practical software development, developers often create commits under conditions of frequent task switching, rapid iteration, and collaborative coordination. However, prior empirical evidence suggests that the quality of commit messages remains a practical concern. Tian et al.~\cite{tian2022goodcommit} found that around 44\% of the commit messages lack either ``what'' or ``why'' information, indicating that many commit messages lack sufficient details on ``what the changes are'' and ``why they were made''. This gap motivates research on the automatic generation of high-quality and semantically informative commit messages, known as commit message generation (CMG), which has emerged as an important research problem in software engineering~\cite{lopes2024llmcommit, li2025cmo}.

Prior studies have explored rule-based, retrieval-based, and learning-based approaches to CMG. With the rapid advancement of large language models (LLMs), researchers have increasingly investigated their application to CMG \cite{lopes2024llmcommit}. To further improve generation quality, subsequent studies have incorporated richer contextual information, including contextual code representations, retrieved historical commit examples, similar diff-commit message pairs, and repository-level code context. However, these contexts remain predominantly focused on code diffs, code context, or repository-internal history. As a result, external information, such as issue reports, remains insufficiently investigated, despite their richer potential to provide the motivation, constraints, and problem context underlying code changes.

In practical software development, developers often report issues on platforms such as GitHub or Jira, providing detailed context, reproduction steps, and constraints. Developers then implement targeted code changes and create commits to address these issues. In this paper, \textbf{issue information} specifically refers to the \textbf{issue title} and \textbf{issue body} associated with a code diff. Such information supplements the code diff with the reported problem context, reproduction steps, and constraints, helping explain why the change was made.

\begin{figure}[htbp]
  \centering
  \includegraphics[width=1\textwidth]{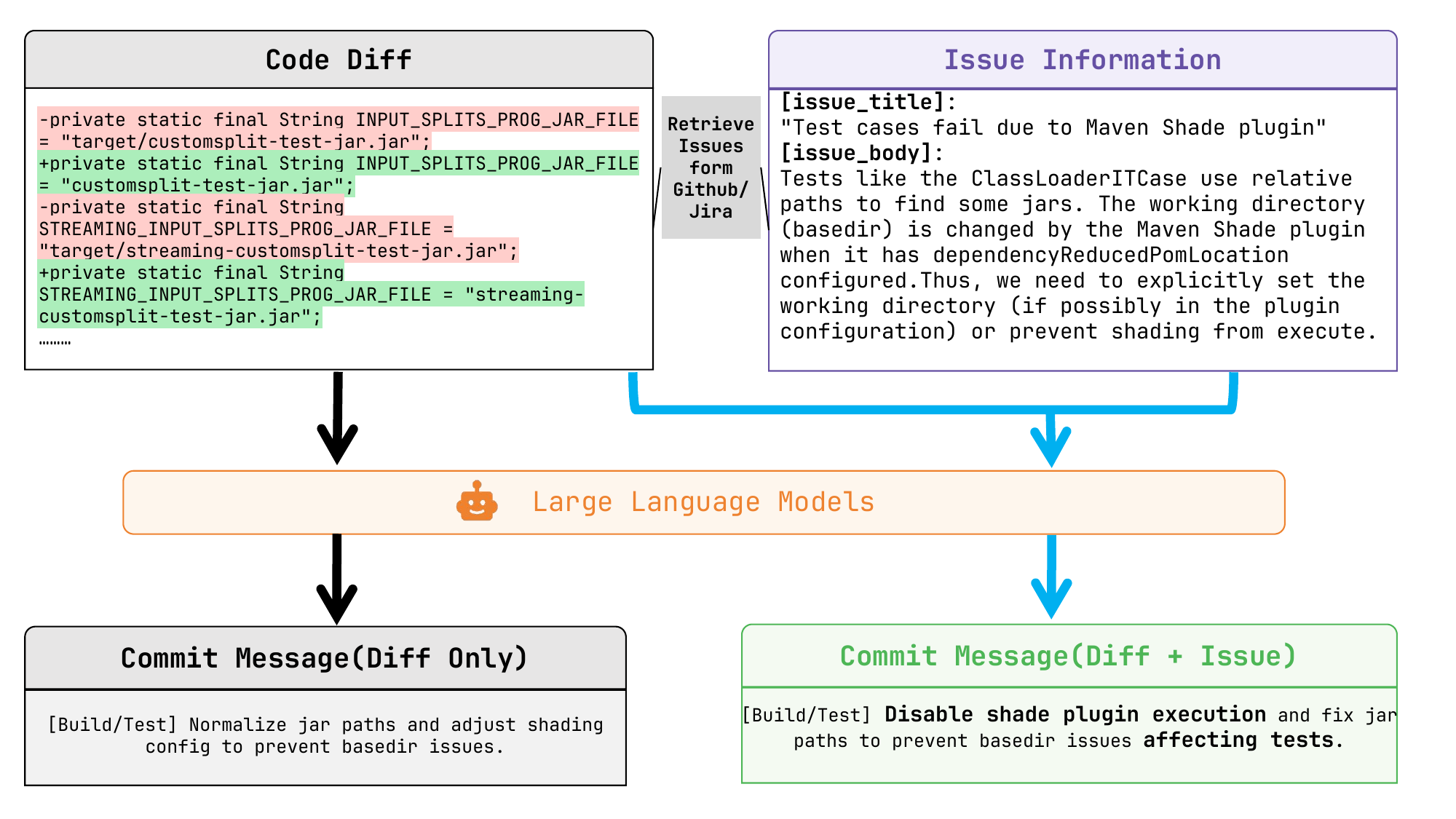}
  \caption{Commit message generated with code diff only \textit{vs.} with code diff and issue information}
  \label{fig:motivation}
\end{figure}

{\textbf{Motivating Example:}} Consider the scenario shown in Figure~\ref{fig:motivation}. For the same code diff, an LLM provided only with the diff produces a relatively generic description that captures the code-level change but fails to convey the underlying problem background or task context. When issue information associated with the code diff is incorporated, the generated commit message becomes more specific: it captures the key code-level operations while also conveying the motivation and impact of the change. As a result, the message is substantially more informative than one generated from the code diff alone. This example illustrates the potential value of issue information for CMG: issue bodies can provide task-level semantics that are difficult to infer from code diffs alone, such as the problem background, design constraints, and expected effects of the change. Motivated by this observation, this paper investigates whether and how issue information can improve LLM-based commit message generation. To this end, we construct a commit--issue aligned dataset and propose an issue-augmented CMG framework that incorporates issue information into the LLM input together with the corresponding code diff.

This paper makes the following \textbf{main contributions}:

(1) \textbf{Construction of the ApacheCM-Issue Dataset}: Based on the ApacheCM dataset~\cite{xiong2025c3gen}, we construct ApacheCM-Issue, a multi-language commit-issue aligned dataset containing code diffs, commit messages, and the associated issue information. The issues are collected from both GitHub and Apache Jira. The dataset provides the data foundation for constructing issue-augmented inputs and also serves as a reusable resource for future research on issue-aware software engineering tasks. The resulting dataset comprises 47,664 issue-commit aligned samples across 49 Apache open-source repositories.

(2) \textbf{An issue-augmented CMG framework}: We propose ISAC, an \textbf{IS}sue-\textbf{A}ugmented framework for \textbf{C}ommit message generation, which incorporates issue information as task-level context together with code diffs as input to LLMs, thereby enhancing LLM-based commit message generation. By enriching code diffs with issue-derived context, ISAC helps the model better capture the rationale underlying code changes.

(3) \textbf{Comprehensive experimental evaluation of the proposed framework}: We evaluate ISAC using DeepSeek-V4-Flash and GPT-5.5, and assess the generated commit messages with five metrics, i.e., BLEU, ROUGE-L, METEOR, CIDEr, and SBERT-Cos. The experiments not only compare the effects of code diffs and issue-augmented inputs, but also analyze the roles of two context organization strategies: similar historical commit examples from the same project and structured issue summaries. The experiment results show that issue information improves the quality of LLM-generated commit messages, similar historical commits provide additional project-specific expression patterns, while structured issue summaries are not suitable as a direct replacement for raw issue text.

\textbf{Paper Organization}: Section~\ref{sec:relatedwork} discusses related studies on CMG. Section~\ref{sec:apachecm-issue-dataset} presents the construction process and demographics of the ApacheCM-Issue dataset. Section~\ref{sec:methodology} introduces the ISAC framework and its three stages. Section~\ref{sec:experimental-setup} describes the four research questions, experimental configurations, LLM selection, and evaluation metrics. Section~\ref{sec:results} reports and analyzes the experimental results from four aspects: issue-text augmentation, comparison with baseline methods, augmentation with similar historical commit examples, and replacement with structured issue summaries. Section~\ref{sec:threats} discusses the threats to the validity of the study, and Section~\ref{sec:conclusion} concludes the paper and outlines future work.
\section{Related Work}
\label{sec:relatedwork}
CMG aims to automatically generate high-quality commit messages from code diffs. Existing CMG approaches can be broadly categorized into rule-based, retrieval-based, learning-based, and LLM-based methods. While early studies primarily generated commit messages from code diffs alone, recent LLM-based approaches have begun to examine how additional software development context can improve the quality of generated commit messages.

\textbf{Rule-based methods} generate commit messages by summarizing code changes with predefined rules, templates, or heuristic strategies. ChangeScribe identifies diff types, impact scopes, and commit features from code diffs, and maps them to predefined templates to produce commit messages~\cite{cortescoy2014changescribe,vasquez2015changescribetool}. Shen et al.~\cite{shen2016automatic} further explored the automatic summarization of both “what” and “why” information from source code changes for commit message generation, highlighting that a complete commit message should cover not only what changed but also why the change was made.

\textbf{Retrieval-based methods} retrieve similar historical code diffs or commit examples from software repositories to help LLMs generate higher-quality commit messages. Context-aware retrieval methods leverage historical commits to capture project-specific terminology and writing styles~\cite{wang2021contextaware}. RACE incorporates retrieved commit examples as exemplars during generation to improve the semantic relevance between generated commit messages and code changes~\cite{shi2022race}. More recent RAG-based approaches combine retrieval with pretrained models and LLMs, showing that retrieved diff--message examples and repository-level code context help generate more accurate, comprehensive, and informative commit messages~\cite{zhang2024ragenhancedcommitmessagegeneration,xiong2025c3gen,xiong2025coracmg}.

\textbf{Learning-based methods} formulate CMG as a neural machine translation task, using large-scale diff--message datasets to train deep neural networks that map code diffs to commit messages~\cite{jiang2017nmtcommit,loyola2017neural}. Subsequent studies, including CoDiSum~\cite{xu2019codisum}, ATOM~\cite{liu2020atom}, CoreGen~\cite{nie2021coregen}, FIRA~\cite{dong2022fira}, COME~\cite{he2023come}, and CCT5~\cite{lin2023cct5}, improve code diff understanding through structural modeling, copy mechanisms, diff representation learning, and pretraining for code diffs. Beyond these diff-only inputs, Wang et al.~\cite{wang2023commitissue} incorporated issue information into CMG by constructing annotated commit--issue datasets and training supervised CMG models with both code diffs and issue-related context. Overall, learning-based methods have advanced the modeling of code diff and improved the quality of generated commit messages; however, they are typically designed as task-specific supervised models that rely on large labeled datasets, and most of them still use the code diff as the primary input.

\textbf{LLM-based methods} frame CMG as a prompting task for general-purpose LLMs, enabling zero-/few-shot generation without training a task-specific model. Zhang et al.~\cite{zhang2024llmcommitstudy} compare Llama~2 and ChatGPT with existing CMG methods and conclude that LLM-generated commit messages have a clear advantage in human evaluation, demonstrating the potential of LLMs for this task. Xue et al.~\cite{xue2024llmcmg} similarly find that LLMs are effective for CMG and that retrieval-based in-context learning can further improve their generation quality. Existing studies further investigate how to construct more effective inputs for LLMs. Context-augmentation approaches provide complementary information beyond the target diff, such as retrieved commit examples, historical project context, or repository-level code context~\cite{xiong2025c3gen,zhang2024ragenhancedcommitmessagegeneration}. Input-refinement approaches instead focus on improving the code diff representation itself by filtering noisy content, selecting salient changes, or condensing large diffs to fit LLM context constraints~\cite{kuang2025brevity}. Li et al.~\cite{Li2024onlydiff} further enhance commit message generation by leveraging ReAct and tool use to incorporate diverse contextual information related to code changes. Although these studies suggest that additional context can benefit LLM-based CMG, relatively few studies have systematically examined how external information, such as issue information, independently affects LLM-generated commit messages.

\textbf{Conclusive Summary}: Existing studies suggest that incorporating additional software development context can improve the quality of LLM-generated commit messages. Prior work has explored various forms of contextual information, including historical commits, repository-level code context, retrieved examples, and issue-related information. However, the specific contribution of issue information itself has not yet been systematically investigated. In particular, issue information can complement code diffs by providing task background, project-specific terminology, and change motivation, but its specific effect on the quality of LLM-generated commit messages remains underexplored.
\section{ApacheCM-Issue Dataset}
\label{sec:apachecm-issue-dataset}



To support the issue-augmented CMG task in this work, we construct a commit-issue aligned dataset named ApacheCM-Issue~\cite{replpack}, based on the high-quality ApacheCM dataset~\cite{xiong2025c3gen}. Specifically, ApacheCM-Issue includes not only the issues, code diffs, and human-written commit messages derived from the original dataset, but also the explicit alignments among these three elements. 
This ApacheCM-Issue dataset not only serves as the experimental foundation for our study (Section~\ref{sec:results}) but also provides a reusable resource for future research on issue-aware software engineering tasks. In this section, we first clarify the motivation and data sources for constructing the ApacheCM-Issue dataset (Section~\ref{motivationanddatasources}), and then provide the detailed procedure for constructing this dataset (Section~\ref{constructionofapacheCMIssue}). 

\subsection{Motivation and Data Source}
\label{motivationanddatasources}

In software development, code changes are typically driven by specific issues, including bug reports, feature requests, or planned improvements. Before implementing these changes, developers usually record the problem description, expected behavior, or implementation goals in the issue trackers. However, this valuable task-level context, which provides essential background and rationale for the change, is often absent from the code diff itself.
Several datasets have been widely used in CMG tasks, including CommitGen~\cite{jiang2017nmtcommit}, NNGen~\cite{liu2018nngen}, CoDiSum~\cite{xu2019codisum}, and MCMD~\cite{tao2021mcmd}. These datasets are collected from open-source GitHub projects and use \texttt{<code diff, commit message>} pairs as the basic data unit, but they lack explicit alignment with the issues that originally motivated the code changes. More recently, although ExGroFi~\cite{wang2023commitissue} constructed a commit-issue parallel dataset, it retains only the basic textual content of commit and issue, omitting valuable issue metadata, such as the issue platform, \texttt{issue\_state}, \texttt{issue\_closed\_at}, \texttt{issue\_created\_at}, and \texttt{issue\_user}. Moreover, ExGroFi focuses only on Java projects, which limits its language diversity and may restrict its applicability in future in-depth experiments on issue-aware CMG across different programming languages. These limitations motivate us to construct ApacheCM-Issue, a dataset that explicitly aligns code diffs with their corresponding issue information, augments the pair with rich issue metadata, and encompasses a wide range of programming languages.

Table~\ref{tab:ch3-datasets} provides a detailed structural comparison between ApacheCM-Issue and the existing datasets. ApacheCM-Issue provides explicit commit--issue alignment, fine-grained issue metadata, and broader programming language coverage. Unlike existing datasets that rely exclusively on GitHub, ApacheCM-Issue combines commits from GitHub with issues sourced from both GitHub Issues and Apache Jira.

\begin{table}[htbp]
  \centering
  \caption{ApacheCM-Issue vs. the existing datasets}
  \label{tab:ch3-datasets}
  \small
  \setlength{\tabcolsep}{5pt}
  \renewcommand{\arraystretch}{1.15}
  \begin{threeparttable}
    \begin{tabular}{@{}ccccc@{}}
      \toprule
      Dataset & Basic unit & Language covered & Issue information & Source platform \\
      \midrule
      CommitGen~\cite{jiang2017nmtcommit}
        & $\langle D,M\rangle$ & Java & No & GitHub \\
      NNGen~\cite{liu2018nngen}
        & $\langle D,M\rangle$ & Java & No & GitHub \\
      CoDiSum~\cite{xu2019codisum}
        & $\langle D,M\rangle$ & Java & No & GitHub \\
      MCMD~\cite{tao2021mcmd}
        & $\langle D,M\rangle$ & Multi-language & No & GitHub \\
      ExGroFi~\cite{wang2023commitissue}
        & $\langle D,I,M\rangle$ & Java & Yes & GitHub \\
      ApacheCM-Issue (Ours)
        & $\langle D,I,M\rangle$ & Multi-language & Yes & GitHub + Jira \\
      \bottomrule
      \multicolumn{5}{@{}l}{Note: $D$ denotes the code diff, $I$ denotes issue information, and $M$ denotes the commit message.}
    \end{tabular}







  \end{threeparttable}
\end{table}

\subsection{Construction of ApacheCM-Issue}\label{constructionofapacheCMIssue}

\subsubsection{Source repositories of issues}
\label{sec:issue source}

The ApacheCM-Issue dataset is built on the ApacheCM dataset~\cite{xiong2025c3gen}, which comprises code diffs and their associated commit messages from high-quality Apache open-source repositories. These ASF (Apache Software Foundation)\footnote{\url{https://www.apache.org/}} projects are community-driven, adhere to standardized commit message conventions with explicit issue references, and maintain complete cross-platform issue tracking across both GitHub Issues and Apache Jira. These characteristics, together with its demonstrated use in recent CMG studies~\cite{xiong2025c3gen, xiong2025coracmg}, establish ApacheCM as an ideal foundation for constructing our commit--issue aligned dataset. 

For this reason, ApacheCM-Issue inherits its commit sources from the ApacheCM project set \cite{xiong2025c3gen}, which encompasses the top 50 high-quality repositories (ranked by star count) under the Apache GitHub organization\footnote{\url{https://github.com/apache}} as its base source repositories, including well-known projects such as Superset, ECharts, Spark, and Flink. However, not every commit can be uniquely matched to an issue, and some projects yield insufficient valid samples after filtering. Consequently, the resulting ApacheCM-Issue dataset contains commit-issue aligned samples derived from 49 of the 50 ApacheCM repositories. The ApacheCM-Issue dataset also provides broad language coverage, directly inherited from its ApacheCM foundation.

\subsubsection{Dataset schema}
To facilitate data retrieval, storage, and experiments in this work, ApacheCM-Issue is organized as a multi-level structured sample indexed by \texttt{commit\_sha}. Each sample comprehensively encapsulates commit metadata, repository metadata, issue metadata, and the code diff. Accordingly, the fields are semantically partitioned into four categories: commit fields, repository fields, linkage fields, and issue fields.

(1) \textbf{Commit fields} directly capture the code change itself, including \codeid{commit\_sha},\allowbreak~\codeid{commit\_message},\allowbreak~\codeid{commit\_author},\allowbreak~\codeid{commit\_date},\allowbreak~and \codeid{diff}.

(2) \textbf{Repository fields} characterize the project and its location associated with a commit, comprising \texttt{repo\_owner}, \texttt{repo\_name}, and \texttt{repo\_url}.

(3) \textbf{Issue fields} are stored in the nested \codeid{issue} object, including \codeid{issue\_number},\allowbreak~\codeid{issue\_title},\allowbreak~\codeid{issue\_state},\allowbreak~\codeid{issue\_created\_at},\allowbreak~\codeid{issue\_closed\_at},\allowbreak~\codeid{issue\_user},\allowbreak~\codeid{issue\_labels},\allowbreak~and \codeid{issue\_body}. Among these fields, \codeid{issue\_title} and \codeid{issue\_body} provide the textual issue content used as the fundamental augmentation for the issue-augmented CMG task.

(4) \textbf{Linkage fields} are used to explicitly associate each commit message with its corresponding external task entry. Specifically, \texttt{issue\_reference} records the issue identifier extracted from the commit message, while the \texttt{issue} object is a nested structured object that encapsulates the complete set of issue-related subfields defined in \textbf{Issue fields}.

\subsubsection{Quality assurance of ApacheCM-Issue}\label{sec:apachecm-issue-quality-control}
Since not every commit in ApacheCM can be reliably linked to a valid issue, we design four filtering rules to exclude low-quality commit-issue aligned pairs, thereby ensuring the overall quality of the ApacheCM-Issue dataset. As listed in Table~\ref{tab:ch3-quality-rules}, the four filtering rules are designed to reduce noisy samples and uphold a clear and reliable alignment between code diffs and their corresponding issues.

\begin{table}[htbp]
  \centering
  \caption{Filtering rules for the ApacheCM-Issue dataset}
  \label{tab:ch3-quality-rules}
  \small
  \setlength{\tabcolsep}{5pt}
  \renewcommand{\arraystretch}{1.18}
  \begin{tabularx}{\textwidth}{@{}l l X@{}}
    \toprule
    No. & Name & Description and rationale \\
    \midrule

     R1 & Explicit Reference Filtering &
    \textbf{Description:} To exclude commits without identifiable issue references based on explicit reference patterns. \newline
    \textbf{Rationale:} To ensure that each retained commit has at least one candidate issue for subsequent retrieval. \\

    \midrule
    R2 & Unique Linkage Filtering &
    \textbf{Description:} To include commits that can be linked to exactly one candidate issue. \newline
    \textbf{Rationale:} To reduce alignment ambiguity and ensure a clearer correspondence between commits and issues.  \\

    \midrule
    R3 & Content Validity Filtering &
    \textbf{Description:} To remove Pull Requests, invalid issue records, and records with insufficient natural-language content. \newline
    \textbf{Rationale:} To ensure that retained issues provide usable natural-language context for issue-augmented CMG. \\

    \midrule
    R4 & Redundancy Filtering &
    \textbf{Description:} Compute the cosine similarity between the issue title and the first line of the commit message with \texttt{all-MiniLM-L6-v2}. Remove pairs with a score above 0.8. \newline
    \textbf{Rationale:} Avoid issue titles that provide an overly direct paraphrase of the reference commit message. \\

    \bottomrule

  \end{tabularx}
\end{table}

To construct the ApacheCM-Issue dataset, we initialize the candidate commit pool with 249,830 base commit records from ApacheCM~\cite{xiong2025c3gen}. Four filtering rules are then applied sequentially to get the commit-issue aligned data for ApacheCM-Issue. More specifically, we first use filtering rule R1 to exclude commits lacking valid issue identifiers and to retain those with explicit issue references. For example, commits containing such references, including GitHub-style identifiers like \texttt{Closes \#123}, \texttt{Fixes \#456}, and \texttt{Resolves \#789}, or Jira-style identifiers like \texttt{SPARK-12345}, are retained as candidate issue-linked commits. After applying R1, 142,339 commits are excluded, leaving 107,491 commits for subsequent filtering with R2. As shown in the second row of Table~\ref{tab:ch3-quality-rules}, R2 removes commits with ambiguous issue linkages and retains only those associated with exactly one candidate issue. This step filters out 1,257 commits, resulting in 106,234 commit-issue pairs. R3 then filters the retrieved candidate issues based on data quality and usability. This step removes 29,750 samples, including GitHub Pull Requests, issues with empty or URL-only content, records with invalid metadata, and samples whose normalized issue titles are identical to the commit message. After applying R3, the dataset contains 76,484 high-quality pairs of real-world commits and their associated issues.

To evaluate issue-augmented CMG tasks, the issue title serves as model input and the commit message as the generation target. High lexical overlap between them can lead to target leakage. Models may exploit this by verbatim or near-verbatim copying, thereby yielding inflated evaluation metrics. To mitigate this threat, we introduce R4 as a target-leakage control step. Specifically, we employ a pre-trained sentence-embedding model \texttt{all-MiniLM-L6-v2} to compute the cosine similarity between the issue title and the first line of the reference commit message. Through preliminary exploration on issue-commit pairs with varying similarity scores, we empirically set the cosine-similarity threshold as 0.8. We observed that pairs exceeding this threshold generally exhibited substantial semantic or lexical overlap, making the issue title a near-direct paraphrase of the reference commit message. For example, an issue titled \texttt{Fix timeout handling} paired with a commit message starting with \texttt{Fix timeout handling in ConnectionManager} would be excluded by R4 because the issue title closely paraphrases the reference message. Finally, R4 removes 28,820 pairs, resulting in a leakage-controlled ApacheCM-Issue containing 47,664 commit-issue aligned samples for the evaluation of the issue-augmented CMG tasks.

Collectively, these filtering rules enable ApacheCM-Issue to maintain a large sample size while substantially reducing noise from missing issue links, ambiguous references, Pull Request contamination, invalid issue information, and exact title duplication. Consequently, each retained sample in ApacheCM-Issue preserves comprehensive structured metadata and provides the essential elements required for issue-augmented CMG tasks, i.e., code diff, human-written commit message, and traceable issue information.

\subsubsection{Demographics of ApacheCM-Issue}

As shown in the first two rows of Table~\ref{tab:ch3-stats}, the ApacheCM-Issue dataset consists of 47,664 commit-issue aligned pairs from 49 Apache repositories. These pairs are further categorized by the primary programming language of each source repository, as summarized in the last three rows of Table~\ref{tab:ch3-stats}. Java dominates with 37,587 pairs (78.9\%), followed by Scala with 6,493 pairs (13.6\%). The remaining 3,584 pairs (7.5\%) originate from repositories primarily written in C++, Rust, TypeScript, Python, Erlang, Go, and Lua. 

Overall, Java and Scala dominate the ApacheCM-Issue dataset, reflecting its heavy concentration on Apache big-data and distributed-system projects, such as Spark, Flink, and Kafka. The inclusion of additional languages, such as C++, Rust, broadens its coverage across diverse programming-language ecosystems, enhancing the generalizability of the findings derived from this dataset. 

\begin{figure}[htbp]
  \centering
  \begin{tikzpicture}
    \begin{axis}[
      ybar,
      width=1.0\linewidth,
      height=6.2cm,
      bar width=4pt,
      ymin=0,
      ymax=70,
      ylabel={Final retention rate (\%)},
      ylabel style={yshift=-4pt},
      xlabel={49 Repositories in the ApacheCM-Issue dataset},
      symbolic x coords={
        airflow, answer, apisix, arrow, beam, brpc, camel, cassandra, couchdb,
        datafusion, dolphinscheduler, doris, druid, dubbo, dubbo-spring-boot-project,
        echarts, flink, flink-cdc, hadoop, hbase, hertzbeat, hive, hudi, iceberg,
        incubator-kie-drools, incubator-seata, incubator-weex, iotdb, jmeter, kafka,
        mesos, mxnet, openwhisk, pinot, predictionio, pulsar, rocketmq, seatunnel,
        shardingsphere, shardingsphere-elasticjob, shenyu, skywalking, spark, storm,
        superset, thrift, tomcat, tvm, zeppelin, zookeeper
      },
      xtick=data,
      enlarge x limits=0.01,
      ymajorgrids=true,
      grid style={dashed,gray!30},
      tick label style={font=\scriptsize},
      label style={font=\small},
      xticklabel style={rotate=90, anchor=east, font=\fontsize{4.5pt}{5pt}\selectfont},
      nodes near coords,
      nodes near coords style={
        font=\fontsize{3.5pt}{4pt}\selectfont,
        rotate=90,
        anchor=west,
        yshift=1pt
      },
      every node near coord/.append style={
        /pgf/number format/fixed,
        /pgf/number format/precision=1
      },
    ]
      \addplot[
        fill=blue!70,
        draw=blue!80!black
      ] coordinates {
        (airflow,0.2467) (answer,2.5641) (apisix,0.2729) (arrow,38.1765)
        (beam,24.8866) (brpc,0.2670) (camel,35.5023) (cassandra,5.3699)
        (couchdb,1.4768) (datafusion,11.1160) (dolphinscheduler,4.1667)
        (doris,0.1621) (druid,0.8923) (dubbo,5.4734)
        (dubbo-spring-boot-project,30.7692) (echarts,14.8982) (flink,41.4926)
        (flink-cdc,13.9384) (hadoop,54.9088) (hbase,43.6760)
        (hertzbeat,0.9901) (hive,26.3008) (hudi,24.1821) (iceberg,0.1428)
        (incubator-kie-drools,0.2806) (incubator-seata,3.3000)
        (incubator-weex,1.9952) (iotdb,12.4728) (jmeter,0.1060)
        (kafka,39.4565) (mesos,0.6932) (mxnet,3.6293) (openwhisk,2.1005)
        (pinot,0.2721) (predictionio,7.1625) (pulsar,2.1842)
        (rocketmq,33.1060) (seatunnel,1.4560) (shardingsphere,17.9118)
        (shardingsphere-elasticjob,2.1834) (shenyu,17.5699)
        (skywalking,2.0274) (spark,29.6087) (storm,33.7188)
        (superset,0.5990) (thrift,45.0130) (tomcat,0.0000) (tvm,0.4918)
        (zeppelin,24.3781) (zookeeper,60.2086)
      };
    \end{axis}
  \end{tikzpicture}
  \caption{Repository-level retention after all filtering rules (The retention rate is the number of final ApacheCM-Issue pairs divided by the number of original ApacheCM commits in the same repository)}
  \label{fig:filtering-retention-by-repository}
\end{figure}
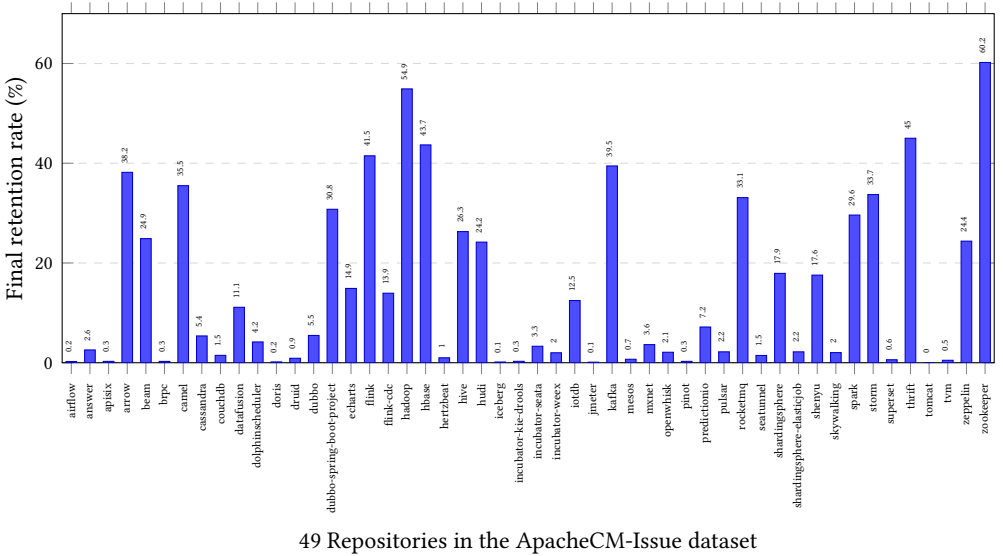

Furthermore, Figure~\ref{fig:filtering-retention-by-repository} illustrates the retention rates at the repository level after applying all four filtering rules to the original commits from ApacheCM. Regarding the 49 included repositories, 47,664 of 249,830 commits are retained, yielding an overall retention rate of 19.08\% in ApacheCM-Issue. As shown in Figure~\ref{fig:filtering-retention-by-repository}, retention proportions vary across the 49 repositories. This indicates that the effectiveness of the four filtering rules is closely dependent on project-specific factors such as commit metadata quality and issue tracking practices.

\begin{table}[htbp]
  \centering
  \caption{Overall statistics of the ApacheCM-Issue dataset}
  \label{tab:ch3-stats}
  \small
  \begin{tabular}{llcl}
    \toprule
    Dimension & Metric & Value & Note \\
    \midrule
    \multirow{2}{*}{Scale}
      & Number of covered repositories & 49 & Apache project repositories \\
    \cmidrule(lr){2-4}
      & Commit--Issue aligned pairs & 47,664 & After all four filtering rules \\
    \midrule
    \multirow{3}{*}{Language}
      & Java & 37,587 (78.9\%) & Primary language of source repository \\
    \cmidrule(lr){2-4}
      & Scala & 6,493 (13.6\%) & Primary language of source repository \\
    \cmidrule(lr){2-4}
      & Other languages & 3,584 (7.5\%) & \makecell[l]{C++, Rust, TypeScript, Python,\\Erlang, Go, and Lua} \\
    \bottomrule
  \end{tabular}
\end{table}

\section{Methodology}\label{sec:methodology}
 
\subsection{ISAC Framework}
\begin{figure}[htbp]
  \centering
  \includegraphics[width=0.95\textwidth]{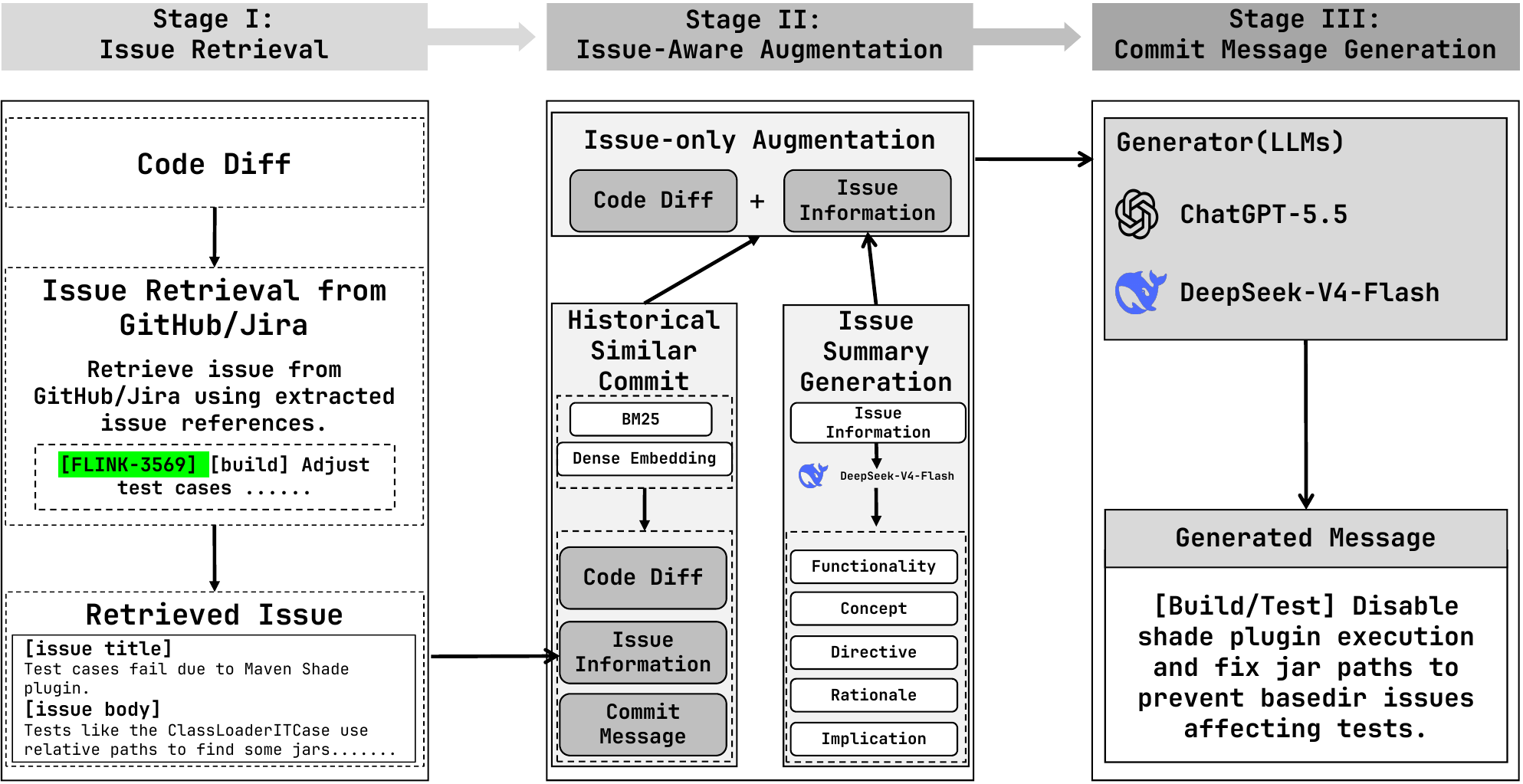}
  \caption{Overview of the ISAC framework}
  \label{fig:methodology-overview}
\end{figure}

Based on the ApacheCM-Issue dataset introduced in
Section~\ref{sec:apachecm-issue-dataset}, we propose the \textbf{IS}sue-\textbf{A}ugmented framework for \textbf{C}ommit message generation (\textbf{ISAC}). As shown in Figure~\ref{fig:methodology-overview}, ISAC encompasses three stages: \textit{Stage I: Issue Retrieval}, \textit{Stage II: Issue-Aware Augmentation}, and \textit{Stage III: Commit Message Generation}. The details of these three stages are provided in the following subsections.


\subsection{Stage I: Issue Retrieval}
Stage I aims to retrieve the issue information associated with the target code diff $d$. The issue records associated with the code diffs have already been retrieved and aligned during the construction of the ApacheCM-Issue dataset. Therefore, Stage I directly obtains the corresponding issue record from ApacheCM-Issue. For each issue record, the \textbf{issue title} and \textbf{issue body} are combined as the issue information, denoted as $i$. We select these two fields because they typically provide the task background, problem context, and implementation intent, which are essential yet rarely inferable from the code diff alone. This stage then produces a diff--issue pair $(d,i)$ as input to the next stage.

\subsection{Stage II: Issue-Aware Augmentation}
\label{sec:StageII}

 In Stage II, we construct the LLM input by further augmenting the diff-issue pair $(d,i)$ produced in Stage I. Specifically, we design three issue augmentation strategies that vary in their formulation of issue information and the additional context incorporated into the prompts.

\begin{figure}[hb]
  \centering
  \includegraphics[width=0.9\textwidth]{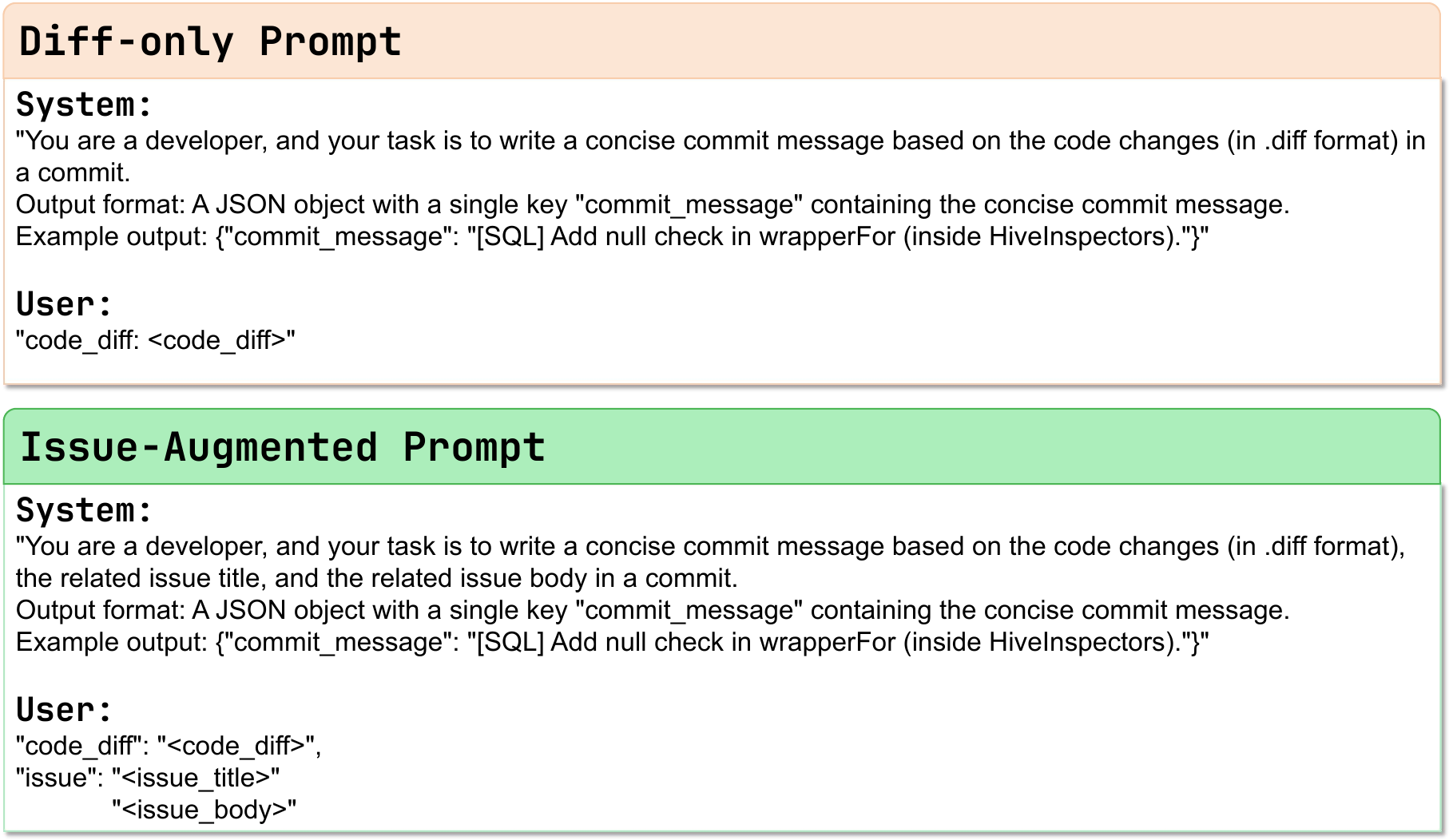}
   \caption{Diff-only prompt template and issue-only augmentation prompt template}
  \label{fig:prompt}
\end{figure}

\textbf{Strategy 1: Issue-only augmentation} integrates the original issue information $i$ with the target code diff $d$. To provide LLMs with a holistic view of the code diff with task-level semantics, we propose a structured prompt template, as illustrated at the bottom of Figure~\ref{fig:prompt}. Different from the Diff-only prompt shown at the top of Figure~\ref{fig:prompt}, our template employs two delimiters, i.e., \texttt{<diff>} and \texttt{<issue>}, to explicitly distinguish concrete code-level modifications from their complementary task-level semantics, including motivation, constraints, and project-specific terminology.

\begin{figure}[hb]
  \centering
  \includegraphics[width=0.9\linewidth]{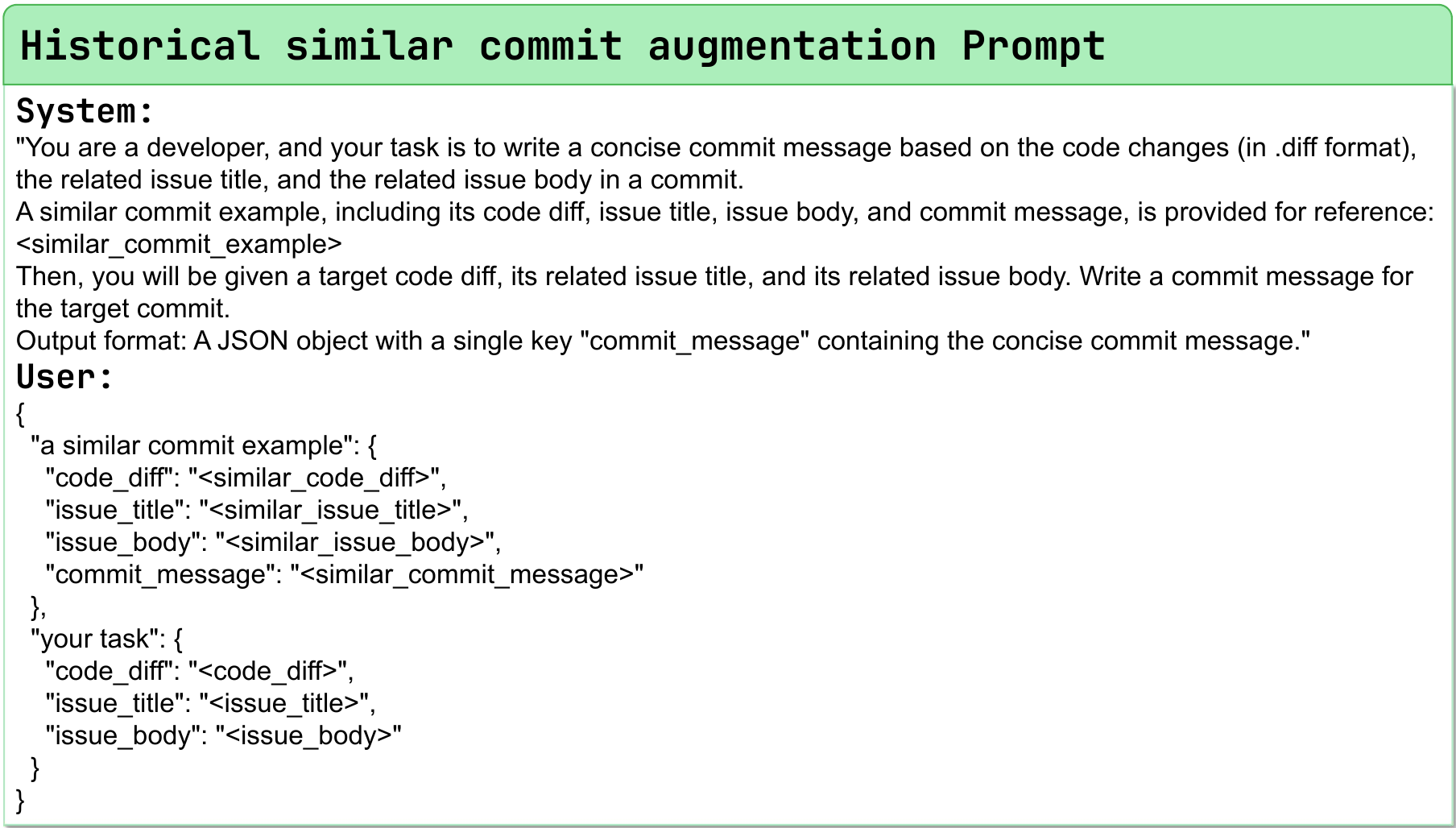}
  \caption{Prompt template for similar historical commit augmentation}
  \label{fig:prompt-history}
\end{figure}

\textbf{Strategy 2: similar historical commit augmentation} extends the original issue-augmented input in Strategy 1 by further incorporating a historically similar commit from the same project into the prompt. Specifically, we searched the ApacheCM-Issue dataset for the most similar historical commit to augment the prompt with its code diff, commit message, and related issue information as a reference example, as shown in Figure~\ref{fig:prompt-history}. The retrieval procedure is detailed in Section~\ref{sec:historical-similar-retrieval}.

\begin{figure}[htbp]
  \centering
  \includegraphics[width=0.9\linewidth]{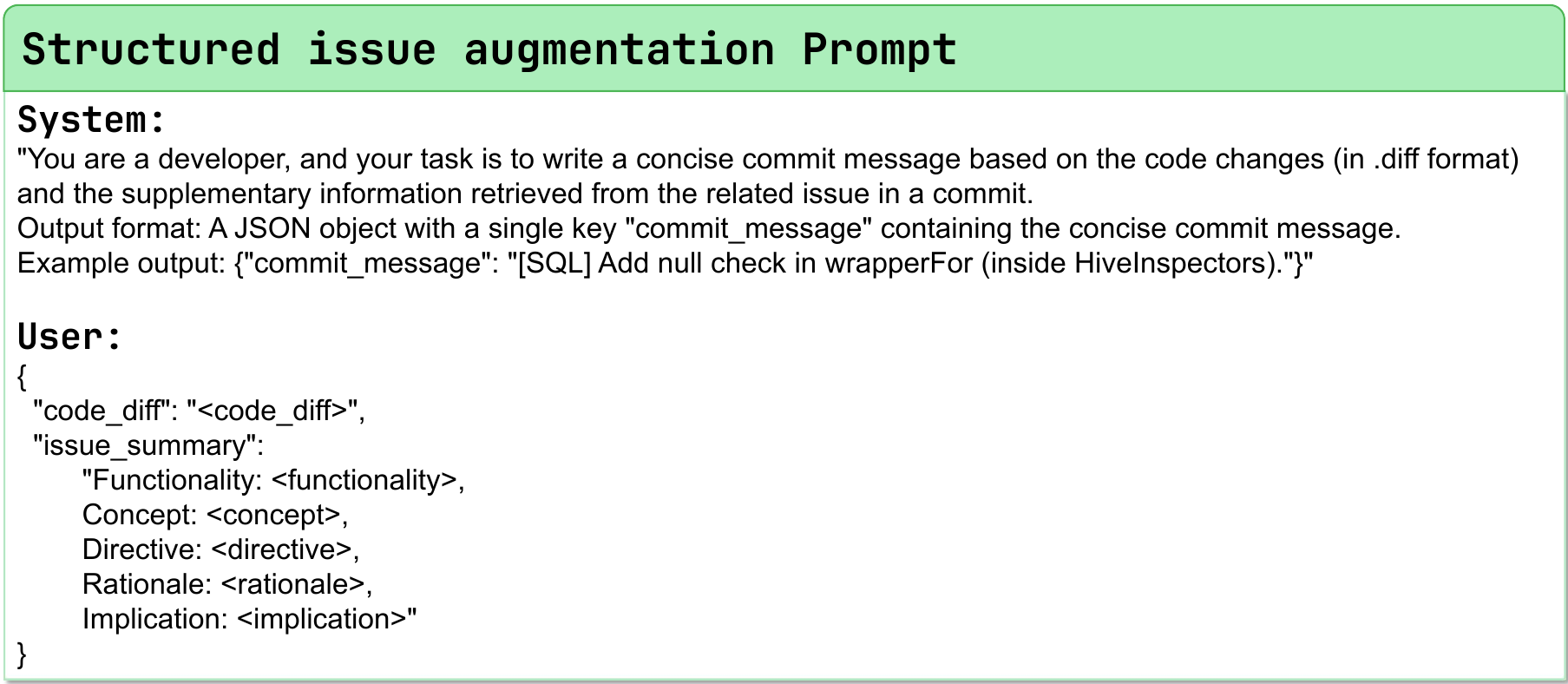}
  \caption{Prompt template for structured issue augmentation}
  \label{fig:prompt-structured}
\end{figure}

\textbf{Strategy 3: Structured issue augmentation} is designed to replace the original issue information used in Strategy 1 with a compressed and structured issue summary. Figure~\ref{fig:prompt-structured} illustrates the prompt template that uses this strategy. 
The procedure for generating the structured issue summary is detailed in Section~\ref{sec:structured-issue-summary}.

Table~\ref{tab:augmentation-strategies} summarizes the information used by these three augmentation strategies.

\begin{table}[htbp]
  \centering
  \caption{Information used by the three augmentation strategies in ISAC}
  \label{tab:augmentation-strategies}
  \small
  \setlength{\tabcolsep}{5pt}
  \renewcommand{\arraystretch}{1.15}
  \begin{tabular}{@{}lcccc@{}}
    \toprule
    Strategy &
    \makecell{Target\\code diff} &
    \makecell{Original\\issue information} &
    \makecell{Structured\\issue summary} &
    \makecell{Historical\\similar commit} \\
    \midrule
    Issue-only augmentation & Yes & Yes & No  & No  \\
    similar historical commit augmentation & Yes & Yes & No  & Yes \\
    Structured issue augmentation   & Yes & No  & Yes & No  \\
    \bottomrule
  \end{tabular}
\end{table}

\subsection{Stage III: Commit Generation}
In Stage III, the augmented input is fed to the LLM to generate a commit message for the target code diff $d$. In this stage, the LLM produces the commit message directly from the structured prompt without any additional fine-tuning. This design isolates the effect of the injected issue information from other factors. To guarantee a consistent and machine-readable output format and facilitate automatic parsing of the generated commit message, the LLM is required to return a JSON object containing exactly one key, \texttt{commit\_message}, as illustrated in the \texttt{<Output>} sections of three prompts in Figure~\ref{fig:prompt}, Figure~\ref{fig:prompt-history}, and Figure~\ref{fig:prompt-structured}.

\subsection{Similar historical commit Retrieval}
\label{sec:historical-similar-retrieval}
To investigate whether historical commits and their associated issue information can further improve issue-augmented CMG, we introduce Strategy~2 in Section~\ref{sec:StageII}. Details of the similar historical commit retrieval process are provided below.

Let $\mathcal{C}= \{c_1,c_2,\ldots,c_N\}$ denote the set of commits in the complete ApacheCM-Issue dataset,
where $N$ is the total number of commit records. Each commit $c_j \in \mathcal{C}$ is represented as a tuple $c_j = \langle d_j,i_j,m_j,r_j,\tau_j \rangle $, where $d_j$, $i_j$, $m_j$, $r_j$, and $\tau_j$ denote the code diff, the associated issue information, the human-written commit message, the repository, and the commit timestamp, respectively.
All these variables are derived from the four field groups defined in Section~\ref{constructionofapacheCMIssue}. Specifically, $d_j$, $m_j$, and $\tau_j$ correspond to \texttt{diff}, \texttt{commit\_message}, and \texttt{commit\_date} in Commit fields; $r_j$ maps to the \texttt{repo\_name} in Repository fields; and $i_j$ represents the associated issue information stored in Issue fields. 



Let $\mathcal{C}_t \subseteq \mathcal{C}$ denote the set of target commits for which commit messages are generated. For a target commit $c_t \in \mathcal{C}_t$, its code diff, issue information, repository, and timestamp are denoted by $d_t$, $i_t$, $r_t$, and $\tau_t$, respectively.
To simulate a realistic development scenario, we restrict the retrieval of historical examples to commits that precede $c_t$ within the same repository. Consequently, the historical candidate set for $c_t$ is defined as:


\begin{equation}
\label{eq:historical-candidate-set}
\mathcal{C}_h(c_t) =
\left\{
c_h \in \mathcal{C}\setminus\{c_t\}
\mid
r_h=r_t \land \tau_h<\tau_t
\right\}.
\end{equation}


For each candidate $c_h \in \mathcal{C}_h(c_t)$, we compare the similarity of its code diff $d_h$ with the target code diff $d_t$ through two complementary metrics. Lexical similarity is computed using BM25, formally defined as: 

\begin{equation}
\label{eq:bm25-score}
LexicalSim(c_t,c_h)=\mathrm{BM25}(d_t,d_h),
\end{equation}

while semantic similarity is computed using their pre-computed dense embeddings and formally defined as:

\begin{equation}
\label{eq:dense-score}
SemanticSim(c_t,c_h)=\mathrm{Dense}(d_t,d_h).
\end{equation}

Due to the disparate value ranges of these two metrics, independent min-max normalization is applied over $\mathcal{C}_h(c_t)$. We denote the normalized scores by $\widehat{LexicalSim}(c_t,c_h)$ and $\widehat{SemanticSim}(c_t,c_h)$, respectively. The final retrieval score is then computed as equal weight to the lexical and semantic similarity.

\begin{equation}
\label{eq:hybrid-retrieval-score}
Sim(c_t,c_h)
=
0.5 \cdot \widehat{LexicalSim}(c_t,c_h)
+
0.5 \cdot \widehat{SemanticSim}(c_t,c_h).
\end{equation}

The candidate with the highest score is selected as the most similar historical commit:

\begin{equation}
\label{eq:top1-historical-commit}
c_s(c_t)
=
\underset{c_h\in\mathcal{C}_h(c_t)}{\arg\max}
\;Sim(c_t,c_h).
\end{equation}

By applying this retrieval procedure to each target commit in $\mathcal{C}_t$, we get the historical retrieval result set as below. 

\begin{equation}
\label{eq:historical-retrieval-result-set}
\mathcal{C}_s
=
\left\{
\langle c_t,c_s(c_t)\rangle
\mid
c_t\in\mathcal{C}_t
\right\}.
\end{equation}

where $d_s$, $i_s$, and $m_s$ denote the code diff, issue information, and commit message of $c_s(c_t)$, respectively. In Stage~II, these fields are added to the target diff and issue information, forming $\langle d_t,i_t,d_s,i_s,m_s\rangle$, to serve as the augmented input for the LLM to generate the commit message for $c_t$ in Stage III.  



\subsection{Structured Issue Summary Generation}\label{sec:structured-issue-summary}
To comprehensively evaluate the impact of different representations of issues on CMG tasks, we propose Strategy 3 (see Section~\ref{sec:StageII}), to examine the effect of replacing the original issue information with a structured issue summary. Generally, original issue information may contain redundant discussions or loosely organized content. For CMG, the most critical information usually includes \textit{what} the change is about, \textit{why} the change is needed, \textit{how} the commit resolves the issue, and what outcome is expected to be delivered. Motivated by this, we constructed a structured issue summary for each issue to replace the original issue information in ISAC, where we employed the LLM demonstrating competitive overall performance to extract the following five fields from the issue information.  

\begin{itemize}
  \item \textbf{Functionality} captures \textit{what} by identifying the functionality, behavior, API, or component that the issue concerns.
  \item \textbf{Concept} complements the \textit{what} by preserving concrete technical entities and project-specific terms, such as configuration names, classes, modules, error messages, or domain concepts.
  \item \textbf{Rationale} captures the \textit{why} by summarizing the reported problem, failure scenario, limitation, or user need that motivates the change.
  \item \textbf{Directive} captures the \textit{how} by describing the modification action through which the commit addresses the issue, such as fixing an inconsistency, supporting a behavior, or improving an implementation.
  \item \textbf{Implication} captures the expected result of the change, including behavior changes, compatibility concerns, testing impact, or performance-related effects.
\end{itemize}

Then, these five fields are organized into a structured JSON object. Under Strategy~3 (structured issue summary augmentation), this object replaces the original issue information in the prompt. This input configuration isolates the effect of issue information representation, enabling us to assess the effectiveness of structured issue summaries for LLM-based CMG.

\section{Research Design}
\label{sec:experimental-setup}


\subsection{Research Questions}
The objective of this study is to investigate how issue information affects the quality of LLM-generated commit messages. To this end, we propose ISAC, an \textbf{IS}sue-\textbf{A}ugmented framework for \textbf{C}ommit message generation, and formulate four research questions (RQs) to examine the impact of issue information on CMG from four perspectives.

\textbf{RQ1: Does ISAC improve the quality of LLM-generated commit messages compared to using only the code diff?} 
Since ISAC is designed to leverage issue information for CMG, this RQ aims to investigate the core motivation of ISAC by comparing the quality of generated commit messages under two input configurations: code diff only versus code diff augmented with issue information. Answering this RQ would help clarify the contribution of issue information to CMG performance.

\textbf{RQ2: Does ISAC outperform state-of-the-art (SOTA) CMG baselines?} 
Following the investigation in RQ1, this RQ is designed to evaluate whether ISAC performs better than existing CMG baselines on the same experimental dataset. The answer to this RQ will help assess the overall effectiveness of ISAC against SOTA methods.

\textbf{RQ3: Can leveraging similar historical commits and their associated issue information from the same project further improve CMG quality when built upon issue information augmentation?} 
This RQ aims to evaluate whether similar historical commits and their associated issue information can further improve the quality of LLM-generated commit messages when issue information is already augmented. We will compare the configuration utilizing only the code diff and issue information with an augmented configuration that additionally incorporates a similar code diff, along with its corresponding issue information and commit message.

\textbf{RQ4: Can structured issue summaries enhance CMG performance compared to using original issue information?} 
This RQ is designed to examine the impact of structured issue summaries on CMG quality across different LLM configurations. Answering this RQ would help understand how structured issue summaries affect CMG performance. 

These four RQs are organized in a progressive manner: \textbf{RQ1} first investigates the core premise of ISAC by examining whether issue information improves CMG quality over code diffs alone. Building on the results of RQ1, \textbf{RQ2} evaluates the overall effectiveness of ISAC by benchmarking it against SOTA CMG baselines. \textbf{RQ3} then explores whether leveraging historically similar commits and their issue information yields incremental improvements beyond basic issue augmentation. Finally, \textbf{RQ4} investigates whether structured issue summaries can serve as an effective alternative to raw issue information in CMG.

\subsection{Experimental Dataset}
Due to cost and time limitations, we construct an experimental dataset based on ApacheCM-Issue~\cite{replpack} to evaluate the performance of ISAC in Section~\ref{sec:results}. Specifically, to assess the generalizability of ISAC across multiple programming languages, we select repositories spanning Scala, Java, and C++. Specifically, Spark and Camel represent Scala and Java, respectively, because both retain sufficient commit-issue pairs after filtering to support random sampling. For C++, no single repository contains enough retained pairs. Therefore, we combine data from all C++ repositories to form the C++ samples.

All candidate pairs have passed the commit-issue alignment, information quality, and target-leakage control filters described in Section~\ref{sec:apachecm-issue-quality-control}. After filtering, the sampling populations comprise 6,439 pairs from Spark (Scala), 6,891 pairs from Camel (Java), and 2,504 pairs from the C++ repositories.

To ensure statistical representativeness, we calculate the required sample size for each language group separately according to Israel~\cite{israel1992determining}. Adopting a 95\% confidence level, a 5\% margin of error, and a population proportion of 0.5 for the most conservative setting, we independently sample 363 pairs from Spark, 364 from Camel, and 334 from the pooled C++ population. In total, the experimental dataset consists of 1,061 commit-issue pairs spanning Scala, Java, and C++.

\subsection{Experiment Configurations}
To systematically evaluate the three issue augmentation strategies outlined in Stage II (see Section~\ref{sec:StageII}), we design the following four distinct input configurations for each sampled commit:

\begin{itemize}
    \item \textsc{Diff}: The control setting, which provides only the code diff to the LLM.
    \item \textsc{Diff + Issue}: Augments the code diff with its related issue information, by implementing Strategy 1 in Section~\ref{sec:StageII}.
    \item \textsc{Diff+Issue+History}: Extends \textsc{Diff+Issue} with the code diff, issue information, and commit message of the top-1 similar historical commit from the same project, thereby realizing Strategy 2 in Section~\ref{sec:StageII}.
    \item \textsc{Diff+Structured Issue}: Combines the target code diff with a structured issue summary that replaces the original issue information, corresponding to Strategy 3 in Section~\ref{sec:StageII}.
\end{itemize}



\subsection{LLM Selection}
To investigate whether the effect of leveraging issue information remains consistent across different LLM configurations, we select four LLM configurations for our experiments, i.e., GPT-5.5 (\texttt{effort=none}), GPT-5.5 (\texttt{effort=high}), DeepSeek-V4-Flash (\texttt{deepseek-chat}), and DeepSeek-V4-Flash (\texttt{deepseek-reasoner}), as listed in Table~\ref{tab:issue-models}. The selected models represent both OpenAI and DeepSeek families, covering non-reasoning and reasoning-enhanced modes. Specifically, we employ DeepSeek-V4-Flash as the dedicated LLM for structured issue summary extraction due to its favorable balance of inference cost, generation speed, and output quality.

\begin{table}[htbp]
  \centering
  \small
  \setlength{\tabcolsep}{5pt}
  \renewcommand{\arraystretch}{1.12}
  \caption{LLMs used in the issue-augmented CMG experiments}
  \label{tab:issue-models}
  \begin{tabular}{@{}lcc@{}}
    \toprule
    LLM configuration & Reasoning configuration \\
    \midrule
    GPT-5.5 (\texttt{effort=none})  & Non-reasoning \\
    GPT-5.5 (\texttt{effort=high})  & Reasoning-enhanced \\
    DeepSeek-V4-Flash (\texttt{deepseek-chat})  & Non-reasoning \\
    DeepSeek-V4-Flash (\texttt{deepseek-reasoner})  & Reasoning-enhanced \\
    \bottomrule
  \end{tabular}
\end{table}

In our experiments, DeepSeek-V4-Flash is accessed via the official DeepSeek API~\cite{deepseek-api-docs} and GPT-5.5 via OpenRouter's OpenAI-compatible API~\cite{openrouter-api-docs}. For each LLM configuration, all generation parameters are held constant across the four input configurations.

\subsection{Evaluation Metrics}
In this paper, the generated commit messages are evaluated with the following five automatic metrics: \textbf{BLEU}~\cite{papineni2002bleu}, \textbf{ROUGE-L}~\cite{lin2004rouge}, \textbf{METEOR}~\cite{banerjee2005meteor}, \textbf{CIDEr}~\cite{vedantam2015cider}, and \textbf{SBERT-Cos}~\cite{reimers2019sentencebert}. Collectively, these metrics can be used to evaluate the quality of LLM-generated commit messages from four perspectives: lexical, structural, term-salience, and semantic.

\begin{itemize}
    \item \textbf{BLEU (lexical)} measures the \textit{n}-gram precision between a generated commit message and the corresponding human-written message, reflecting how many word fragments in the generated message are consistent with the reference.

    \item \textbf{ROUGE-L (structural)} evaluates sequence-level overlap based on the longest common subsequence and therefore enables capturing whether the generated message preserves the main word order and structure of the reference message.

    \item \textbf{METEOR (lexical)} considers unigram-level matching with both precision and recall, and gives relatively higher importance to recall, making it more sensitive to whether important content in the reference commit message is covered.

    \item \textbf{CIDEr (term-salience)} represents each message using TF-IDF weighted \textit{n}-grams, assigning larger weights to informative and project-specific terms, which helps distinguish generic overlap from more meaningful technical alignment.

    \item \textbf{SBERT-Cos (semantic)} computes the cosine similarity between sentence embeddings produced by Sentence-BERT, thereby measuring semantic closeness even when the generated and reference messages use different surface forms.
\end{itemize}

\subsection{Human Evaluation}
\label{sec:human-evaluation}

To complement the automatic evaluation, we conduct a human evaluation of the generated commit messages. Using stratified random sampling over the full 1,061 experimental samples, we select 50 samples for human evaluation. The resulting evaluation set contains 17 Scala samples, 17 Java samples, and 16 C++ samples. 

For each sample, human evaluators are provided with the target code diff, the related issue information, and 17 anonymized candidate commit messages. Among these 17 candidates, 16 are generated under the four LLM configurations and four experimental settings, while the remaining one is the original human-written commit message in the dataset. The identities and generation settings of the candidates are hidden from the evaluators. To prevent project and issue identifiers from affecting the assessment, we also remove leading issue-related prefixes, such as \texttt{[SPARK-XXXXX]}, from all candidate messages.
We invited two evaluators to conduct the human evaluation: an expert and a graduate student specializing in software engineering. Both evaluators have substantial software development experience and are familiar with version control, code review, and commit message conventions. They independently assess each candidate commit message in terms of \textbf{Clarity}, \textbf{Completeness}, and \textbf{Correctness}. More specifically, \textbf{Clarity} measures whether the message is precise, concise, and easy to understand. \textbf{Completeness} assesses whether it covers the main code changes and their purpose. \textbf{Correctness} evaluates whether the statements are consistent with the code diff and the issue information. Each criterion is rated on a five-point Likert scale, as defined in Table~\ref{tab:human-evaluation-criteria}. For each evaluator, we compute the average score of each criterion across the 50 samples for each experimental configuration.

\begin{table}[htbp]
  \centering
  \caption{Criteria used in the human evaluation of generated commit messages}
  \label{tab:human-evaluation-criteria}
  \small
  \setlength{\tabcolsep}{4pt}
  \renewcommand{\arraystretch}{1.15}
  \begin{tabularx}{\textwidth}{@{}cXXX@{}}
    \toprule
    Score & Clarity & Completeness & Correctness \\
    \midrule
    1 & Difficult to understand and poorly expressed.
      & Describes almost none of the main changes.
      & Clearly inconsistent with the actual change. \\
    2 & Contains substantial ambiguity or unnatural wording.
      & Omits most important changes.
      & Contains major factual errors. \\
    3 & Generally understandable but contains some ambiguity or redundancy.
      & Describes the main change but omits important information.
      & Generally correct but contains minor errors or unsupported information. \\
    4 & Clear, with only minor expression problems.
      & Covers the main change and purpose, with only minor omissions.
      & Consistent with the input and contains no substantial errors. \\
    5 & Precise, concise, and easy to understand.
      & Fully covers the main change and its purpose.
      & Every statement is supported by the code diff or issue information. \\
    \bottomrule
  \end{tabularx}
\end{table}

\section{Results and Analysis}\label{sec:results}
This section evaluates the proposed ISAC framework from the following four perspectives: (1) the impact of augmenting code diffs with issue information in CMG, (2) the performance of ISAC against the replicated baselines, (3) the additional effect of incorporating similar historical commits, and (4) the effectiveness of replacing the original issue information with structured issue summaries.


\subsection{Effectiveness of ISAC in Leveraging Issue Information (RQ1)}

\subsubsection{Results of RQ1}
Table~\ref{tab:isac-performance} compares \textsc{Diff+Issue}, which applies issue-only augmentation (i.e.,Strategy 1) of ISAC, with the \textsc{Diff} configuration across different LLM configurations. It is observed that augmenting code diffs with issue information consistently improves all five metrics for every LLM configuration. Compared to the baseline using only code diffs, integrating only issue information in ISAC yields average improvements ranging from 8.70\% in SBERT-Cos to 30.70\% in CIDEr. BLEU, ROUGE-L, and METEOR exhibit noteworthy average gains of 23.60\%, 20.48\%, and 25.21\%, respectively. Moreover, the magnitude of the improvement varies across LLM configurations and metrics. DeepSeek-V4-Flash (\texttt{chat}) achieves the largest relative gains on BLEU, ROUGE-L, and CIDEr, while DeepSeek-V4-Flash (\texttt{reasoner}) achieves the largest gains on METEOR and SBERT-Cos. In absolute terms, GPT-5.5 (\texttt{effort=high}) achieves the highest BLEU, ROUGE-L, CIDEr, and SBERT-Cos scores under \textsc{Diff+Issue}, while GPT-5.5 (\texttt{effort=none}) achieves the highest METEOR score.


\begin{table}[H]
  \centering
  \footnotesize
  \setlength{\tabcolsep}{3.5pt}
  \renewcommand{\arraystretch}{1.15}
  \caption{Performance comparison between ISAC and the \textsc{Diff} setting across different LLM configurations}
  \label{tab:isac-performance}
  \resizebox{\textwidth}{!}{
  \begin{tabular}{@{}llccccc@{}}
    \toprule
    Model & \makecell[c]{Input\\configuration}
    & BLEU & ROUGE-L & METEOR & CIDEr & SBERT-Cos \\
    \midrule
    \multirow{2}{*}{\makecell[l]{DeepSeek-V4-Flash\\(\texttt{reasoner})}}
      & \textsc{Diff}
      & 13.53 & 25.76 & 25.08 & 9.68 & 56.37 \\
    & \textsc{Diff+Issue}
      & \makecell{\textbf{17.00}\\($\uparrow$ 25.64\%)}
      & \makecell{\textbf{31.29}\\($\uparrow$ 21.49\%)}
      & \makecell{\textbf{33.44}\\($\uparrow$ 33.33\%)}
      & \makecell{\textbf{12.76}\\($\uparrow$ 31.78\%)}
      & \makecell{\textbf{62.32}\\($\uparrow$ 10.56\%)} \\

    \midrule
    \multirow{2}{*}{\makecell[l]{DeepSeek-V4-Flash\\(\texttt{chat})}}
      & \textsc{Diff}
      & 13.58 & 25.81 & 25.65 & 9.86 & 56.91 \\
    & \textsc{Diff+Issue}
      & \makecell{\textbf{17.53}\\($\uparrow$ 29.08\%)}
      & \makecell{\textbf{31.73}\\($\uparrow$ 22.92\%)}
      & \makecell{\textbf{33.59}\\($\uparrow$ 30.94\%)}
      & \makecell{\textbf{13.38}\\($\uparrow$ 35.77\%)}
      & \makecell{\textbf{62.43}\\($\uparrow$ 9.69\%)} \\

    \midrule
    \multirow{2}{*}{\makecell[l]{GPT-5.5\\(\texttt{effort=high})}}
      & \textsc{Diff}
      & 15.62 & 27.81 & 28.98 & 11.05 & 59.69 \\
    & \textsc{Diff+Issue}
      & \makecell{\textbf{18.68}\\($\uparrow$ 19.58\%)}
      & \makecell{\textbf{32.79}\\($\uparrow$ 17.90\%)}
      & \makecell{\textbf{34.17}\\($\uparrow$ 17.92\%)}
      & \makecell{\textbf{14.06}\\($\uparrow$ 27.27\%)}
      & \makecell{\textbf{63.89}\\($\uparrow$ 7.03\%)} \\

    \midrule
    \multirow{2}{*}{\makecell[l]{GPT-5.5\\(\texttt{effort=none})}}
      & \textsc{Diff}
      & 15.34 & 27.40 & 28.91 & 10.81 & 59.41 \\
    & \textsc{Diff+Issue}
      & \makecell{\textbf{18.43}\\($\uparrow$ 20.10\%)}
      & \makecell{\textbf{32.76}\\($\uparrow$ 19.59\%)}
      & \makecell{\textbf{34.30}\\($\uparrow$ 18.64\%)}
      & \makecell{\textbf{13.83}\\($\uparrow$ 27.97\%)}
      & \makecell{\textbf{63.88}\\($\uparrow$ 7.52\%)} \\

    \midrule
    \multicolumn{2}{l}{\textbf{Average change}}
      & \textbf{$\uparrow$ 23.60\%}
      & \textbf{$\uparrow$ 20.48\%}
      & \textbf{$\uparrow$ 25.21\%}
      & \textbf{$\uparrow$ 30.70\%}
      & \textbf{$\uparrow$ 8.70\%} \\
    \bottomrule
  \end{tabular}}
\end{table}

\begin{figure}[H]
\centering
\begin{tikzpicture}
\begin{axis}[
    ybar,
    width=0.88\linewidth,
    height=5.4cm,
    bar width=10pt,
    ylabel={Relative improvement over \textsc{Diff} (\%)},
    ylabel style={yshift=-4pt},
    symbolic x coords={BLEU,ROUGE-L,METEOR,CIDEr,SBERT-Cos},
    xtick=data,
    ymin=0,
    ymax=45,
    enlarge x limits=0.10,
    ymajorgrids=true,
    grid style={dashed,gray!30},
    legend style={
        at={(0.5,1.02)},
        anchor=south,
        legend columns=2,
        draw=none,
        font=\small
    },
    tick label style={font=\small},
    label style={font=\small},
    nodes near coords,
    nodes near coords style={
        font=\fontsize{4pt}{4.5pt}\selectfont,
        rotate=0,
        anchor=south,
        yshift=1pt
    },
    every node near coord/.append style={
        /pgf/number format/fixed,
        /pgf/number format/precision=2
    },
]

\addplot[
    fill=blue!70,
    draw=blue!80!black
] coordinates {
    (BLEU,25.64) (ROUGE-L,21.49) (METEOR,33.33) (CIDEr,31.78) (SBERT-Cos,10.56)
};

\addplot[
    fill=blue!35,
    draw=blue!60!black
] coordinates {
    (BLEU,29.08) (ROUGE-L,22.92) (METEOR,30.94) (CIDEr,35.77) (SBERT-Cos,9.69)
};

\addplot[
    fill=orange!80,
    draw=orange!90!black
] coordinates {
    (BLEU,19.58) (ROUGE-L,17.90) (METEOR,17.92) (CIDEr,27.27) (SBERT-Cos,7.03)
};

\addplot[
    fill=orange!40,
    draw=orange!70!black
] coordinates {
    (BLEU,20.10) (ROUGE-L,19.59) (METEOR,18.64) (CIDEr,27.97) (SBERT-Cos,7.52)
};

\legend{
    DeepSeek-V4-Flash reasoner,
    DeepSeek-V4-Flash chat,
    GPT-5.5 effort=high,
    GPT-5.5 effort=none
}
\end{axis}
\end{tikzpicture}
\caption{Relative improvement achieved by ISAC over the \textsc{Diff} setting across different LLM configurations.}
\label{fig:issue-improvement}
\end{figure}
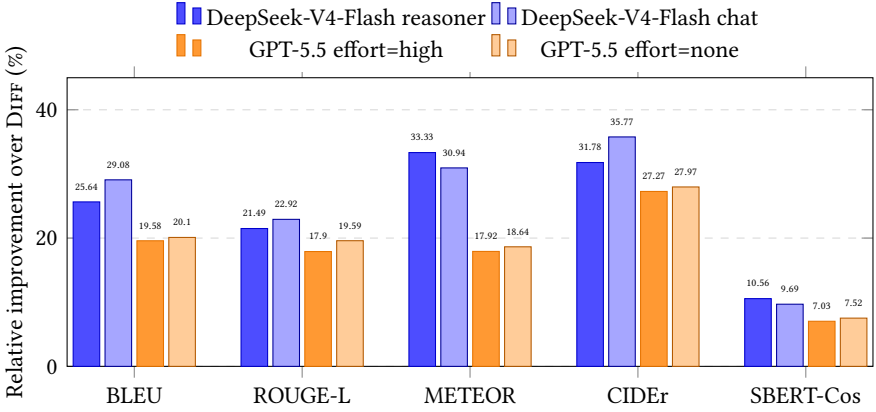

Figure~\ref{fig:issue-improvement} illustrates the relative improvements achieved by incorporating issue information, which vary across LLM configurations and metrics. Notably, the DeepSeek family consistently outperforms the OpenAI family when augmenting issue information with code diff in CMG, especially in the average improvement in METEOR. By contrast, both families exhibit comparable improvements in ROUGE-L and SBERT-Cos.

\subsubsection{Analysis of RQ1 Results}\leavevmode\\

\noindent\textbf{Impact of issue augmentation.} 
When comparing the \textsc{Diff} and \textsc{Diff+Issue} settings, the issue-augmented input contributes higher scores across all four LLM configurations and all five metrics. On average, adding issue information improves BLEU, ROUGE-L, METEOR, CIDEr, and SBERT-Cos by 23.60\%, 20.48\%, 25.21\%, 30.70\%, and 8.70\%, respectively. These results indicate that the issue information provides supplementary context beyond what is available from code diffs alone. While code diffs mainly describe how the source code is changed, issue information often explains why the change is needed, such as the reported defect, requested feature, expected behavior, or reported problem. This additional information enables LLMs to generate commit messages that better reflect the intent behind the change, rather than merely offering a superficial summary of the code diffs. 

Considering the five evaluation metrics, CIDEr shows the highest average improvement, indicating that issue augmentation is especially beneficial for producing task-specific and reference-aligned expressions. This is reasonable because issue information often contains key domain terms, problem descriptions, and requirement-related phrases. In contrast, SBERT-Cos shows a smaller but still consistent improvement, suggesting that issue information improves semantic alignment. Overall, the average gains demonstrate that incorporating issue information improves lexical and phrase-level alignment with reference commit messages, as well as semantic relevance in LLM-based CMG.\\

\noindent\textbf{Performance of ISAC with different LLM configurations.}
Issue information benefits both model families and both reasoning settings, albeit with varying degrees of relative improvement. Among all evaluated configurations, DeepSeek-V4-Flash exhibits the largest relative improvement on all five metrics. In the \textsc{Diff+Issue} setting, GPT-5.5 (\texttt{effort=high}) achieves the best performance across BLEU, ROUGE-L, CIDEr, and SBERT-Cos scores, scoring 18.68, 32.79, 14.06, and 63.89, respectively; while GPT-5.5 (\texttt{effort=none}) achieves the highest METEOR score of 34.30.
Reasoning-enhanced configurations do not consistently outperform their non-reasoning counterparts. Specifically, GPT-5.5 (\texttt{effort=high}) performs slightly better than \texttt{effort=none} on BLEU, ROUGE-L, CIDEr, and SBERT-Cos, but slightly worse on METEOR. For DeepSeek-V4-Flash, \texttt{chat} obtains higher scores than \texttt{reasoner} on all five automatic metrics under \textsc{Diff+Issue}. These results indicate that the benefit of issue information is robust across inference settings, but increasing reasoning effort alone does not lead to universally higher automatic metric scores for CMG.

\begin{keyfindingbox}{Key Findings of RQ1:}
\begin{itemize}
  \item Leveraging issue information enables ISAC to consistently outperform the \textsc{Diff} configuration across all four evaluated LLM configurations and five automatic metrics.
  \item The improvement brought by issue information is most significant on CIDEr, indicating that ISAC better captures task-specific and reference-aligned expressions for commit messages.
  \item Among the evaluated configurations, GPT-5.5 (\texttt{effort=high}) demonstrates the best overall performance, achieving the highest scores on four of the five automatic metrics. In contrast, DeepSeek-V4-Flash (\texttt{reasoner}) exhibits the weakest performance, with the lowest scores on all five automatic metrics.
\end{itemize}
\end{keyfindingbox}

\subsection{Effectiveness of ISAC Compared with SOTA CMG Baselines (RQ2)}

\subsubsection{Results of RQ2}
To further validate the effectiveness of ISAC, we compare Strategy~1 (issue-only augmentation) with four replicated SOTA CMG baselines on the same experimental dataset. We focus on Strategy~1 because it represents the core design of ISAC: augmenting a code diff directly with its original issue information. To cover the best- and worst-performing configurations of Strategy~1, we select two representative configurations from Table~\ref{tab:isac-performance}. DeepSeek-V4-Flash (\texttt{reasoner}), which obtains the lowest scores across all five metrics under \textsc{Diff+Issue}, is denoted as $ISAC_{worst}$. GPT-5.5 (\texttt{effort=high}), which achieves the strongest overall performance by ranking first on four of the five metrics and nearly first on METEOR, is denoted as $ISAC_{best}$.

As shown in Table~\ref{tab:baseline-performance}, both the worst-performing and best-performing ISAC configurations outperform all the replicated SOTA CMG baselines across all five automatic metrics. This result demonstrates the effectiveness of ISAC over existing CMG methods on our experimental dataset.

\begin{table}[htbp]
  \centering
  \footnotesize
  \setlength{\tabcolsep}{5pt}
  \renewcommand{\arraystretch}{1.15}
  \caption{Performance comparison between SOTA CMG baselines and ISAC configurations}
  \label{tab:baseline-performance}
  \begin{tabular}{@{}lccccc@{}}
    \toprule
    Model & BLEU & ROUGE-L & METEOR & CIDEr & SBERT-Cos \\
    \midrule
    CCT5~\cite{lin2023cct5}  & 13.78 & 22.94 & 22.16 & 10.96 & 47.06 \\
    NNGen~\cite{liu2018nngen} & 8.73 & 12.52 & 16.82 & 3.16 & 29.94 \\
    RACE~\cite{shi2022race}  & 9.27 & 17.08 & 15.75 & 3.40 & 36.95 \\
    CoRec~\cite{wang2021contextaware} & 10.87 & 14.65 & 19.39 & 4.26 & 32.78 \\
    \midrule
    \makecell[l]{$ISAC_{worst}$\\(DeepSeek-V4-Flash \texttt{reasoner})}
      & 17.00 & 31.29 & 33.44 & 12.76 & 62.32 \\
    \makecell[l]{$ISAC_{best}$\\(GPT-5.5 \texttt{effort=high})}
      & 18.68 & 32.79 & 34.17 & 14.06 & 63.89 \\
    \bottomrule
  \end{tabular}
\end{table}

\subsubsection{Analysis of RQ2 Results}\leavevmode\\

\noindent\textbf{Comparison with reproduced baselines.}
To establish the baseline comparison, we reproduce four SOTA CMG baselines: CCT5~\cite{lin2023cct5}, NNGen~\cite{liu2018nngen}, RACE~\cite{shi2022race}, and CoRec~\cite{wang2021contextaware}. Among them, CCT5 is the strongest reproduced baseline, achieving the best results across all five metrics. Compared with CCT5, even $ISAC_{\mathrm{worst}}$, represented by DeepSeek-V4-Flash (\texttt{reasoner}), yields improvements of 23.37\%, 36.40\%, 50.90\%, 16.42\%, and 32.43\% on BLEU, ROUGE-L, METEOR, CIDEr, and SBERT-Cos, respectively. Furthermore, $ISAC_{\mathrm{best}}$, represented by GPT-5.5 (\texttt{effort=high}), increases the corresponding improvements to 35.56\%, 42.94\%, 54.20\%, 28.28\%, and 35.76\%, respectively. The consistent improvements across all automatic metrics demonstrate that ISAC reliably outperforms the strongest reproduced SOTA CMG baseline, even under its least effective LLM configuration.

These improvements suggest that ISAC enables LLM-based CMG to generate commit messages that are more consistent with reference commit messages than reproduced SOTA CMG baselines. The improvements on BLEU, ROUGE-L, METEOR, and CIDEr reflect stronger lexical and phrase-level alignment between generated commit messages and reference commit messages, while the improvement on SBERT-Cos indicates better semantic relevance between generated commit messages and reference commit messages.

\begin{keyfindingbox}{Key Findings of RQ2:}
\begin{itemize}
    \item By merely augmenting code diffs with original issue information, ISAC consistently outperforms the four replicated SOTA CMG baselines across all five automatic metrics, even under its worst-performing LLM configuration.
\end{itemize}
\end{keyfindingbox}

\subsection{Effect of Similar Historical Commit Augmentation (RQ3)}

\subsubsection{Results of RQ3}
Table~\ref{tab:historical-augmentation-performance} compares the performance of ISAC under the \textsc{Diff+Issue} and \textsc{Diff+Issue+History} settings. The latter configuration adds the code diff, issue information, and commit message of the most similar historical commit from the same project. Across all four LLM configurations, incorporating the most similar historical commit consistently improves the performance of ISAC on every evaluation metric, with average relative gains of 29.49\%, 14.84\%, 13.67\%, 41.08\%, and 2.66\% for BLEU, ROUGE-L, METEOR, CIDEr, and SBERT-Cos, respectively.

\begin{table}[htbp]
  \centering
  \footnotesize
  \setlength{\tabcolsep}{3.5pt}
  \renewcommand{\arraystretch}{1.15}
  \caption{Effect of incorporating a similar historical commit on top of original issue augmentation}
  \label{tab:historical-augmentation-performance}
  \resizebox{\textwidth}{!}{
  \begin{tabular}{@{}llccccc@{}}
    \toprule
    Model & \makecell[c]{Input\\configuration} & BLEU & ROUGE-L & METEOR & CIDEr & SBERT-Cos \\
    \midrule
    \multirow{2}{*}{\makecell[l]{DeepSeek-V4-Flash\\(\texttt{reasoner})}}
      & \textsc{Diff+Issue} & 17.00 & 31.29 & 33.44 & 12.76 & 62.32 \\
      & \textsc{Diff+Issue+History}
      & \makecell{\textbf{22.19}\\($\uparrow$ 30.55\%)}
      & \makecell{\textbf{35.72}\\($\uparrow$ 14.14\%)}
      & \makecell{\textbf{38.00}\\($\uparrow$ 13.65\%)}
      & \makecell{\textbf{18.24}\\($\uparrow$ 42.97\%)}
      & \makecell{\textbf{63.71}\\($\uparrow$ 2.23\%)} \\
    \midrule
    \multirow{2}{*}{\makecell[l]{DeepSeek-V4-Flash\\(\texttt{chat})}}
      & \textsc{Diff+Issue} & 17.53 & 31.73 & 33.59 & 13.38 & 62.43 \\
      & \textsc{Diff+Issue+History}
      & \makecell{\textbf{22.44}\\($\uparrow$ 28.02\%)}
      & \makecell{\textbf{36.35}\\($\uparrow$ 14.55\%)}
      & \makecell{\textbf{37.48}\\($\uparrow$ 11.60\%)}
      & \makecell{\textbf{18.98}\\($\uparrow$ 41.80\%)}
      & \makecell{\textbf{63.67}\\($\uparrow$ 2.00\%)} \\
    \midrule
    \multirow{2}{*}{\makecell[l]{GPT-5.5\\(\texttt{effort=high})}}
      & \textsc{Diff+Issue} & 18.68 & 32.79 & 34.17 & 14.06 & 63.89 \\
      & \textsc{Diff+Issue+History}
      & \makecell{\textbf{24.52}\\($\uparrow$ 31.31\%)}
      & \makecell{\textbf{38.01}\\($\uparrow$ 15.93\%)}
      & \makecell{\textbf{39.50}\\($\uparrow$ 15.60\%)}
      & \makecell{\textbf{19.85}\\($\uparrow$ 41.20\%)}
      & \makecell{\textbf{66.29}\\($\uparrow$ 3.76\%)} \\
    \midrule
    \multirow{2}{*}{\makecell[l]{GPT-5.5\\(\texttt{effort=none})}}
      & \textsc{Diff+Issue} & 18.43 & 32.76 & 34.30 & 13.83 & 63.88 \\
      & \textsc{Diff+Issue+History}
      & \makecell{\textbf{23.61}\\($\uparrow$ 28.11\%)}
      & \makecell{\textbf{37.60}\\($\uparrow$ 14.75\%)}
      & \makecell{\textbf{39.05}\\($\uparrow$ 13.85\%)}
      & \makecell{\textbf{19.14}\\($\uparrow$ 38.37\%)}
      & \makecell{\textbf{65.56}\\($\uparrow$ 2.64\%)} \\
    \midrule
    \multicolumn{2}{l}{\textbf{Average change}}
      & \textbf{$\uparrow$ 29.49\%}
      & \textbf{$\uparrow$ 14.84\%}
      & \textbf{$\uparrow$ 13.67\%}
      & \textbf{$\uparrow$ 41.08\%}
      & \textbf{$\uparrow$ 2.66\%} \\
    \bottomrule
  \end{tabular}}
\end{table}

Among all four LLM configurations, GPT-5.5 (\texttt{effort=high}) achieves the highest absolute performance under \textsc{Diff+Issue+History}, with absolute scores of 24.52, 38.01, 39.50, 19.85, and 66.29 on BLEU, ROUGE-L, METEOR, CIDEr, and SBERT-Cos, respectively. In particular, this LLM configuration also exhibits the largest relative gain in BLEU (31.31\%). Meanwhile, the CIDEr metric shows remarkable consistency across all four LLM configurations, with gains tightly bounded between 38.37\% and 42.97\%. This underscores the robust and consistent benefit of leveraging similar historical commits.

\subsubsection{Analysis of RQ3 Results}\leavevmode\\
\noindent\textbf{Benefits of Incorporating Similar Historical Commits for CMG.}
Incorporating similar historical commits consistently improves the performance of ISAC across all five automatic metrics and LLM configurations. 
On average, CIDEr and BLEU exhibit the most pronounced gains, rising by 41.08\% and 29.49\%, respectively, followed by ROUGE-L (14.84\%), METEOR (13.67\%), and SBERT-Cos (2.66\%). 
This benefit generalizes across all four LLM configurations. Specifically, GPT-5.5 (\texttt{effort=high}) achieves the highest absolute scores and the largest relative improvements on BLEU, ROUGE-L, METEOR, and SBERT-Cos, while DeepSeek-V4-Flash (\texttt{reasoner}) obtains the highest CIDEr gain of 42.97\%. These results indicate that augmenting issue information with its similar historical commit yields superior quality of generated commit messages beyond issue information augmentation alone.

\begin{keyfindingbox}{Key Findings of RQ3:}
\begin{itemize}
  \item Incorporating similar historical commits and their related issue information consistently improves all five evaluation metrics across the four LLM configurations.
\end{itemize}
\end{keyfindingbox}

\subsection{Effect of Structured Issue Augmentation (RQ4)}

\subsubsection{Results of RQ4}
Table~\ref{tab:structured-issue-performance} compares the performance of ISAC under \textsc{Diff+Issue} and \textsc{Diff+Structured Issue}. 
It is observed that in all four LLM configurations, replacing the original issue title and body with structured issue summaries consistently leads to performance degradation across all five evaluation metrics, since each LLM configuration achieves lower scores under \textsc{Diff+Structured Issue} than under \textsc{Diff+Issue}.
On average across four LLM configurations, BLEU, ROUGE-L, METEOR, CIDEr, and SBERT-Cos decrease by 9.90\%, 6.99\%, 9.86\%, 10.65\%, and 3.46\%, respectively.

\begin{table}[htbp]
  \centering
  \footnotesize
  \setlength{\tabcolsep}{3.5pt}
  \renewcommand{\arraystretch}{1.15}
  \caption{Performance comparison between original issue information and structured issue summary augmentation}
  \label{tab:structured-issue-performance}
  \resizebox{\textwidth}{!}{
  \begin{tabular}{@{}llccccc@{}}
    \toprule
    Model & \makecell[c]{Input\\configuration} & BLEU & ROUGE-L & METEOR & CIDEr & SBERT-Cos \\
    \midrule
    \multirow{2}{*}{\makecell[l]{DeepSeek-V4-Flash\\(\texttt{reasoner})}}
      & \textsc{Diff+Issue} & \textbf{17.00} & \textbf{31.29} & \textbf{33.44} & \textbf{12.76} & \textbf{62.32} \\
      & \textsc{Diff+Structured Issue}
      & \makecell{15.12\\($\downarrow$ 11.05\%)}
      & \makecell{28.89\\($\downarrow$ 7.69\%)}
      & \makecell{28.91\\($\downarrow$ 13.55\%)}
      & \makecell{11.49\\($\downarrow$ 9.99\%)}
      & \makecell{59.75\\($\downarrow$ 4.12\%)} \\
    \midrule
    \multirow{2}{*}{\makecell[l]{DeepSeek-V4-Flash\\(\texttt{chat})}}
      & \textsc{Diff+Issue} & \textbf{17.53} & \textbf{31.73} & \textbf{33.59} & \textbf{13.38} & \textbf{62.43} \\
      & \textsc{Diff+Structured Issue}
      & \makecell{14.98\\($\downarrow$ 14.53\%)}
      & \makecell{29.18\\($\downarrow$ 8.04\%)}
      & \makecell{28.79\\($\downarrow$ 14.28\%)}
      & \makecell{11.79\\($\downarrow$ 11.89\%)}
      & \makecell{59.64\\($\downarrow$ 4.46\%)} \\
    \midrule
    \multirow{2}{*}{\makecell[l]{GPT-5.5\\(\texttt{effort=high})}}
      & \textsc{Diff+Issue} & \textbf{18.68} & \textbf{32.79} & \textbf{34.17} & \textbf{14.06} & \textbf{63.89} \\
      & \textsc{Diff+Structured Issue}
      & \makecell{17.52\\($\downarrow$ 6.20\%)}
      & \makecell{31.12\\($\downarrow$ 5.09\%)}
      & \makecell{32.63\\($\downarrow$ 4.51\%)}
      & \makecell{12.71\\($\downarrow$ 9.61\%)}
      & \makecell{62.46\\($\downarrow$ 2.23\%)} \\
    \midrule
    \multirow{2}{*}{\makecell[l]{GPT-5.5\\(\texttt{effort=none})}}
      & \textsc{Diff+Issue} & \textbf{18.43} & \textbf{32.76} & \textbf{34.30} & \textbf{13.83} & \textbf{63.88} \\
      & \textsc{Diff+Structured Issue}
      & \makecell{16.99\\($\downarrow$ 7.82\%)}
      & \makecell{30.43\\($\downarrow$ 7.12\%)}
      & \makecell{31.86\\($\downarrow$ 7.11\%)}
      & \makecell{12.30\\($\downarrow$ 11.11\%)}
      & \makecell{61.96\\($\downarrow$ 3.00\%)} \\
    \midrule
    \multicolumn{2}{l}{\textbf{Average change}}
      & \textbf{$\downarrow$ 9.90\%}
      & \textbf{$\downarrow$ 6.99\%}
      & \textbf{$\downarrow$ 9.86\%}
      & \textbf{$\downarrow$ 10.65\%}
      & \textbf{$\downarrow$ 3.46\%} \\
    \bottomrule
  \end{tabular}}
\end{table}

Among the four LLM configurations, GPT-5.5 (\texttt{effort=high}) is least affected by the replacement of structured issue summaries, with performance degrading by 6.20\%, 5.09\%, 4.51\%, 9.61\%, and 2.23\% across the five metrics. In contrast, DeepSeek-V4-Flash (\texttt{chat}) shows the largest decreases in BLEU, ROUGE-L, METEOR, and SBERT-Cos. 

\subsubsection{Analysis of RQ4 Results}\leavevmode\\
\noindent\textbf{Information loss introduced by structured issue summaries.}
As shown in Table~\ref{tab:structured-issue-performance}, although the structured issue summary condenses each issue into five predefined fields, this compression consistently reduces ISAC's performance compared to using the original issue information. The largest average decreases are observed in CIDEr, BLEU, and METEOR, which are more sensitive to informative terms and content coverage. This suggests that the summarization process tends to omit concrete details that are valuable for CMG, such as exact API names, configuration keys, error messages, affected modules, conditions, and expected behaviors. While the structured format preserves a general description of what, why, and how of an issue, it fails to retain the full range of project-specific expressions in the original issue.

Meanwhile, the smaller decrease in SBERT-Cos indicates that structured issue summaries preserve much of the broad semantic intent of the original issue. However, as the preceding analysis indicates, retaining general intent alone is insufficient to match the specificity required by human-written commit messages. Therefore, the original issue information offers a more effective balance between intent-level semantics and concrete technical evidence essential for CMG.\\ 

\noindent\textbf{Effect of the reasoning setting of LLMs}
GPT-5.5 (\texttt{effort=high}) exhibits greater robustness to issue compression than the other configurations, particularly on METEOR and SBERT-Cos. This suggests that increased reasoning effort may help recover relational information among the structured fields of the issue summary. Nevertheless, it does not reverse the overall performance degradation, as all five evaluation metrics remain below the scores achieved with the original issue. The consistent degradation is replicated across both model families (ChatGPT and DeepSeek) suggests that the limitation is independent of specific LLM configurations. Instead, directly replacing the source issue with a fixed structured summary removes information that all evaluated configurations can exploit.

\begin{keyfindingbox}{Key Findings of RQ4:}
\begin{itemize}
  \item Replacing the original issue with a structured issue summary decreases all five automatic metrics across all four LLM configurations.
  \item While structured issue summaries retain the broad semantics of original issue information, they may omit technical details and project-specific expressions that are essential for generating high-quality commit messages.
\end{itemize}
\end{keyfindingbox}

\subsection{Human Evaluation}
\label{sec:human-evaluation-results}

\subsubsection{Results of Human Evaluation}
Following the human evaluation protocol described in Section~\ref{sec:experimental-setup}, we collected individual criterion scores from both evaluators for the 50 sampled cases under each experimental configuration.
Table~\ref{tab:human-evaluation-performance} shows the average scores over the two evaluators.

\begin{table}[htbp]
  \centering
  \footnotesize
  \setlength{\tabcolsep}{4pt}
  \renewcommand{\arraystretch}{1.08}
  \caption{Human evaluation results averaged across two evaluators (1--5 scale)}
  \label{tab:human-evaluation-performance}
  \resizebox{\textwidth}{!}{
  \begin{tabular}{@{}llccc@{}}
    \toprule
   LLM configuration & Experiment configuration & Clarity & Completeness & Correctness \\
    \midrule
    \multirow{4}{*}{\makecell[l]{DeepSeek-V4-Flash\\(\texttt{chat})}}
      & \textsc{Diff} & 4.34 & 3.49 & 4.55 \\
      & \textsc{Diff+Issue} & 4.53 & 3.84 & 4.60 \\
      & \textsc{Diff+Issue+History} & 4.34 & 3.41 & 4.50 \\
      & \textsc{Diff+Structured Issue} & 4.39 & 3.86 & 4.46 \\
    \midrule
    \multirow{4}{*}{\makecell[l]{DeepSeek-V4-Flash\\(\texttt{reasoner})}}
      & \textsc{Diff} & 4.36 & 3.65 & 4.61 \\
      & \textsc{Diff+Issue} & 4.59 & 3.87 & 4.70 \\
      & \textsc{Diff+Issue+History} & 4.29 & 3.32 & 4.43 \\
      & \textsc{Diff+Structured Issue} & 4.52 & 3.80 & 4.72 \\
    \midrule
    \multirow{4}{*}{\makecell[l]{GPT-5.5\\(\texttt{effort=none})}}
      & \textsc{Diff} & 4.40 & 3.50 & 4.65 \\
      & \textsc{Diff+Issue} & 4.42 & 3.44 & 4.58 \\
      & \textsc{Diff+Issue+History} & 4.20 & 3.06 & 4.32 \\
      & \textsc{Diff+Structured Issue} & 4.45 & 3.59 & 4.59 \\
    \midrule
    \multirow{4}{*}{\makecell[l]{GPT-5.5\\(\texttt{effort=high})}}
      & \textsc{Diff} & 4.33 & 3.37 & 4.59 \\
      & \textsc{Diff+Issue} & 4.53 & 3.49 & 4.63 \\
      & \textsc{Diff+Issue+History} & 4.23 & 3.13 & 4.38 \\
      & \textsc{Diff+Structured Issue} & 4.45 & 3.56 & 4.65 \\
    \midrule
    Human-written & Original Commit Message & 3.68 & 3.18 & 4.01 \\
    \bottomrule
  \end{tabular}}
\end{table}

\subsubsection{Analysis of Human Evaluation Results}
Table~\ref{tab:human-evaluation-performance} reports the average ratings from the two evaluators. Compared with \textsc{Diff}, the \textsc{Diff+Issue} setting improves clarity for all four LLM configurations and enhances the completeness of the generated commit messages in most cases. This suggests that issue information enables LLMs to articulate not only the change itself but also its underlying rationale. 
Correctness remains comparable to the \textsc{Diff} setting, indicating that these improvements do not compromise the consistency of the generated commit messages with the code diff and issue information.

The \textsc{Diff+Issue+History} setting consistently underperforms across all three human evaluation scores. While the retrieved examples provide useful project context and examples of historical issue-related commits, they may also introduce potentially irrelevant information. This additional context may cause LLMs to either incorporate unrelated details or overlook important changes in the target code diff.

The \textsc{Diff+Structured Issue} configuration exhibits a different pattern. It achieves the highest average completeness, suggesting that organizing issue content into explicit fields facilitates LLMs in covering the core elements of a change. However, the clarity and correctness of structured issue summaries do not surpass those produced with the original issue information. This indicates that while the structured issue summaries improve the completeness of generated commit messages, the compression of issues may remove contextual details from the original issue information that are necessary to describe the change precisely.

The lower human-evaluation ratings for human-written messages suggest that such messages may not always serve as ideal references to assess the clarity and completeness of generated commit messages. Reference-based automatic metrics primarily measure how closely a generated commit message aligns with the reference message, rather than directly evaluating its quality. This difference in evaluation focus may explain why automatic and human evaluations do not always exhibit consistent trends across experimental configurations.

\begin{keyfindingbox}{Key Findings of the Human Evaluation:}
\begin{itemize}
  \item Original issue information improves clarity and completeness without compromising correctness of generated commit messages. LLM-generated commit messages generally receive higher ratings than human-written commit messages.
  \item Human evaluation and automatic metrics do not always show consistent trends. Human-written commit messages may not always provide ideal references for evaluating generated commit messages.
\end{itemize}
\end{keyfindingbox}


\section{Threats to Validity}
\label{sec:threats}
In this section, we discuss the threats to the validity of our study from four perspectives: internal validity, external validity, construct validity, and conclusion validity.

\subsection{Internal Validity}
Internal validity concerns potential factors within the experimental design that may affect the interpretation of the observed improvements. In this study, the main internal threat comes from the quality of commit-issue alignment. As described in Section~\ref{sec:apachecm-issue-dataset}, we applied multiple filtering strategies, including unique candidate constraints, pull request filtering, and content quality filtering, to improve the reliability of the aligned pairs. However, imperfect alignments cannot be fully eliminated. For example, a commit may include changes beyond the problem described in the linked issue, or it may only resolve part of the linked issue. In such cases, the issue information may not accurately reflect the background of the code diff, the purpose of the modification, or the actual scope of the change, which could affect our estimation of the actual benefit of issue information for LLM-based commit message generation.
Another internal threat lies in the varying informativeness of issue content. Some issues contain clear problem descriptions, reproduction steps, functional requirements, or project-specific terminology, whereas others may include only short titles, generic descriptions, or fragmented discussions. Even when the commit-issue alignment is correct, such differences in issue content quality may influence how much useful context the issue provides to the LLM for CMG. Such variation in issue informativeness is difficult to fully avoid and may introduce fluctuations in LLM performance across different commits.

\subsection{External Validity}
External validity concerns the generalizability of our methods to other projects, ecosystems, and development workflows. ApacheCM-Issue \cite{replpack} contains 47,664 commit--issue aligned samples from 49 Apache repositories, and the 1,061 experimental samples are drawn from six projects: Camel, Spark, Arrow, Thrift, MXNet, and Mesos. This multi-project experimental dataset reduces the dependence of our findings on a single repository, but the dataset remains concentrated in established Apache open-source projects and is not evenly distributed across projects. The study results may therefore not transfer directly to proprietary systems, small-scale projects, or projects governed by different communities, where commit-message conventions, issue-tracking practices, and issue quality may differ substantially.

Our findings also apply primarily to commits with an identifiable and usable issue record. The dataset construction process requires an explicit commit--issue link and applies several quality and leakage-control filters. Commits without external issue context, or commits whose linked issue provides little relevant information, are therefore underrepresented in the experimental data. In addition, the retrieval of similar historical commits is restricted to earlier commits from the same project, and consequently its effectiveness in improving LLM-based CMG may differ in projects with shorter histories, sparse issue tracking, or different repository structures. Finally, we evaluated four configurations from two LLM families, which do not cover the full range of available models and inference settings. Evaluating ISAC on more diverse projects, development environments, and LLM configurations remains necessary to further assess its generalizability.

\subsection{Construct Validity}
Construct validity concerns whether the experimental design and measurements properly capture the concepts that this study aims to evaluate. In this work, we use \texttt{issue title} and \texttt{issue body} as the source of issue information for LLM-based CMG. However, issue information may also include other fields, such as the issue state, creation and closing time, labels, comments, and developer information. By focusing only on \texttt{issue title} and \texttt{issue body}, our experiments may not fully capture all contextual information contained in an issue.

We evaluated generated commit messages using BLEU, ROUGE-L, METEOR, CIDEr, and SBERT-Cos. These automatic metrics measure similarity to human-written reference commit messages from different perspectives, but they remain approximate proxies for indicating commit message quality. We therefore complemented them with a human evaluation of the clarity, completeness, and correctness of generated commit messages. The two forms of evaluation do not always show the same trends because they capture different aspects of commit message quality. Moreover, the human evaluation is based on 50 samples and two evaluators, and the ratings may not represent the preferences of the broader developer population.

\subsection{Conclusion Validity}
Threats to conclusion validity concern whether the observed experimental results provide reliable support for our conclusions. To improve the transparency and reproducibility of the experimental process, we implemented dataset construction, issue retrieval, prompt construction, model invocation, and result aggregation as scripted pipelines, and released the complete experimental source code~\cite{replpack}. The released code includes the fixed experimental samples, input templates, model configurations, and evaluation procedures, enabling other researchers to replicate the overall experimental workflow. However, the evaluated LLMs were accessed through online API services, and their providers may update the underlying model implementations without exposing all internal changes. Consequently, rerunning the experiments at a later time may produce slightly different numerical results even when the same code and experimental settings are used. To mitigate this threat, we base our conclusions primarily on consistent relative performance differences across multiple model configurations and evaluation metrics, rather than on individual absolute scores.

\section{Conclusions \& Future Work}\label{sec:conclusion}
This paper proposed ISAC, an issue-augmented framework for LLM-based commit message generation, and constructed the ApacheCM-Issue dataset \cite{replpack} to provide aligned code diffs, commit messages, and issue information. Across all evaluated model configurations, adding original issue information improves all five automatic metrics and outperforms the reproduced SOTA CMG baselines. Similar historical commits further improve the reference-based metrics, while structured issue summaries perform worse than the original issue text in the automatic metrics. The human evaluation also shows that original issue information improves clarity and generally enhances completeness without compromising the correctness of generated commit messages. Structured issue summaries achieve the highest completeness ratings, whereas retrieved historical examples may introduce information unrelated to the current change. Overall, the study results show that issue information helps LLMs capture the intent behind code changes, while the way this context is selected and presented affects the quality of the generated messages.

Future work can extend this study in three directions. First, more fine-grained selection strategies could identify the issue fragments and similar historical commits most relevant to the target code diff, thereby reducing noise from weakly related context. Second, structured representations could be refined to better preserve details relevant to CMG or combined with the original issue information, rather than directly replacing it as input to LLM-based CMG. Third, human evaluation could be extended to more projects, samples, and professional developers to assess the usefulness and acceptability of issue-augmented commit messages in practice.

\section*{Data Availability}
The code and data used in this study are publicly available in our GitHub repository~\cite{replpack}. The repository includes the constructed Commit-Issue aligned dataset, data-processing scripts, prompts, model outputs, evaluation scripts, and documentation required to reproduce the analysis.

\begin{acks}
This work has been partially supported by the National Natural Science Foundation of China (NSFC) with Grant No. 92582203. 
\end{acks}

\bibliographystyle{ACM-Reference-Format}
\bibliography{references}

@inproceedings{jiang2017nmtcommit,
  author={Jiang, Siyuan and Armaly, Ameer and McMillan, Collin},
  booktitle={Proceedings of the 32nd IEEE/ACM International Conference on Automated Software Engineering (ASE)}, 
  title={Automatically generating commit messages from diffs using neural machine translation}, 
  year={2017},
  pages={135--146}
}

@inproceedings{xu2019codisum,
  title     = {Commit Message Generation for Source Code Changes},
  author    = {Xu, Shengbin and Yao, Yuan and Xu, Feng and Gu, Tianxiao and Tong, Hanghang and Lu, Jian},
  booktitle = {Proceedings of the 28th International Joint Conference on Artificial Intelligence (IJCAI)},
  pages     = {3975--3981},
  year      = {2019}
}

@inproceedings{shi2022race,
    title = "RACE: Retrieval-augmented Commit Message Generation",
    author = "Shi, Ensheng and Wang, Yanlin and Tao, Wei and Du, Lun and Zhang, Hongyu and Han, Shi and Zhang, Dongmei and Sun, Hongbin",
    booktitle = "Proceedings of the 27th Conference on Empirical Methods in Natural Language Processing (EMNLP)",
    year = "2022",
    pages = "5520--5530"
}

@article{lopes2024llmcommit,
  title   = {Commit Messages in the Age of Large Language Models},
  author  = {Lopes, Cristina V. and Klotzman, Vanessa I. and Ma, Iris and Ahmed, Iftekar},
  journal = {arXiv preprint arXiv:2401.17622},
  year    = {2024}
}

@inproceedings{cortescoy2014changescribe,
  author={Cortés-Coy, Luis Fernando and Linares-Vásquez, Mario and Aponte, Jairo and Poshyvanyk, Denys},
  booktitle={Proceedings of the 14th IEEE International Working Conference on Source Code Analysis and Manipulation (SCAM)}, 
  title={On Automatically Generating Commit Messages via Summarization of Source Code Changes}, 
  year={2014},
  pages={275--284}
}

@inproceedings{vasquez2015changescribetool,
  author={Linares-Vásquez, Mario and Cortés-Coy, Luis Fernando and Aponte, Jairo and Poshyvanyk, Denys},
  booktitle={Proceedings of the 37th IEEE/ACM IEEE International Conference on Software Engineering (ICSE)}, 
  title={ChangeScribe: A Tool for Automatically Generating Commit Messages}, 
  year={2015},
  pages={709--712}
}

@inproceedings{loyola2017neural,
    title = "A Neural Architecture for Generating Natural Language Descriptions from Source Code Changes",
    author = "Loyola, Pablo and Marrese-Taylor, Edison and Matsuo, Yutaka",
    booktitle = "Proceedings of the 55th Annual Meeting of the Association for Computational Linguistics (ACL): Short Papers",
    year = "2017",
    pages = "287--292"
}

@article{wang2021contextaware,
author = {Wang, Haoye and Xia, Xin and Lo, David and He, Qiang and Wang, Xinyu and Grundy, John},
title = {Context-aware Retrieval-based Deep Commit Message Generation},
year = {2021},
volume = {30},
number = {4},
journal = {ACM Transactions on Software Engineering Methodolology},
pages = {1--30}
}

@article{liu2020atom,
  author={Liu, Shangqing and Gao, Cuiyun and Chen, Sen and Nie, Lun Yiu and Liu, Yang},
  journal={IEEE Transactions on Software Engineering}, 
  title={ATOM: Commit Message Generation Based on Abstract Syntax Tree and Hybrid Ranking}, 
  year={2022},
  volume={48},
  number={5},
  pages={1800--1817},
}

@article{nie2021coregen,
  author = {Nie, Lun Yiu and Gao, Cuiyun and Zhong, Zhicong and Lam, Wai and Liu, Yang and Xu, Zenglin},
  title = {CoreGen: Contextualized Code Representation Learning for Commit Message Generation},
  year = {2021},
  volume = {459},
  journal = {Neurocomputing},
  pages = {97--107}
}

@inproceedings{dong2022fira,
  author={Dong, Jinhao and Lou, Yiling and Zhu, Qihao and Sun, Zeyu and Li, Zhilin and Zhang, Wenjie and Hao, Dan},
  booktitle={Proceedings of the 44th IEEE/ACM International Conference on Software Engineering (ICSE)}, 
  title={FIRA: Fine-Grained Graph-Based Code Change Representation for Automated Commit Message Generation}, 
  year={2022},
  pages={970--981}
}

@inproceedings{wang2023commitissue,
  author={Wang, Liran and Tang, Xunzhu and He, Yichen and Ren, Changyu and Shi, Shuhua and Yan, Chaoran and Li, Zhoujun},
  booktitle={Proceedings of the 38th IEEE/ACM International Conference on Automated Software Engineering (ASE)}, 
  title={Delving into Commit-Issue Correlation to Enhance Commit Message Generation Models}, 
  year={2023},
  pages={710--722}
}

@article{li2025cmo,
      title={Consider What Humans Consider: Optimizing Commit Message Leveraging Contexts Considered By Human}, 
      author={Jiawei Li and David Faragó and Christian Petrov and Iftekhar Ahmed},
      year={2025},
      journal={arXiv preprint arXiv:2503.11960},
}

@inproceedings{liu2018nngen,
  author={Liu, Zhongxin and Xia, Xin and Hassan, Ahmed E. and Lo, David and Xing, Zhenchang and Wang, Xinyu},
  booktitle={Proceedings of the 33rd IEEE/ACM International Conference on Automated Software Engineering (ASE)}, 
  title={Neural-Machine-Translation-Based Commit Message Generation: How Far Are We?}, 
  year={2018},
  pages={373--384}
}

@inproceedings{tao2021mcmd,
  author={Tao, Wei and Wang, Yanlin and Shi, Ensheng and Du, Lun and Han, Shi and Zhang, Hongyu and Zhang, Dongmei and Zhang, Wenqiang},
  booktitle={Proceedings of the 37th IEEE International Conference on Software Maintenance and Evolution (ICSME)}, 
  title={On the Evaluation of Commit Message Generation Models: An Experimental Study}, 
  year={2021},
  pages={126--136}
}

@inproceedings{papineni2002bleu,
    title = "BLEU: a Method for Automatic Evaluation of Machine Translation",
    author = "Papineni, Kishore and Roukos, Salim  and Ward, Todd  and Zhu, Wei-Jing",
    booktitle = "Proceedings of the 40th Annual Meeting of the Association for Computational Linguistics (ACL)",
    year = "2002",
    pages = "311--318"
}

@inproceedings{lin2004rouge,
    title = "ROUGE: A Package for Automatic Evaluation of Summaries",
    author = "Lin, Chin-Yew",
    booktitle = "Text Summarization Branches Out",
    year = "2004",
    pages = "74--81"
}

@inproceedings{banerjee2005meteor,
    title = "METEOR: An Automatic Metric for MT Evaluation with Improved Correlation with Human Judgments",
    author = "Banerjee, Satanjeev  and Lavie, Alon",
    booktitle = "Proceedings of the ACL Workshop on Intrinsic and Extrinsic Evaluation Measures for Machine Translation and/or Summarization",
    year = "2005",
    pages = "65--72"
}

@INPROCEEDINGS{vedantam2015cider,
  author={Vedantam, Ramakrishna and Zitnick, C. Lawrence and Parikh, Devi},
  booktitle={Proceedings of the 33rd IEEE Conference on Computer Vision and Pattern Recognition (CVPR)}, 
  title={CIDEr: Consensus-based image description evaluation},
  year={2015},
  pages={4566--4575}
}

@inproceedings{reimers2019sentencebert,
    title = "Sentence-BERT: Sentence Embeddings using Siamese BERT-Networks",
    author = "Reimers, Nils  and Gurevych, Iryna",
    booktitle = "Proceedings of the 2019 Conference on Empirical Methods in Natural Language Processing and the 9th International Joint Conference on Natural Language Processing (EMNLP-IJCNLP)",
    year = "2019",
    pages = "3982--3992"
}

@inproceedings{he2023come,
  author = {He, Yichen and Wang, Liran and Wang, Kaiyi and Zhang, Yupeng and Zhang, Hang and Li, Zhoujun},
  title = {COME: Commit Message Generation with Modification Embedding},
  year = {2023},
  booktitle = {Proceedings of the 32nd ACM SIGSOFT International Symposium on Software Testing and Analysis},
  pages = {792--803}
}

@inproceedings{lin2023cct5,
  author = {Lin, Bo and Wang, Shangwen and Liu, Zhongxin and Liu, Yepang and Xia, Xin and Mao, Xiaoguang},
  title = {CCT5: A Code-Change-Oriented Pre-trained Model},
  year = {2023},
  booktitle = {Proceedings of the 31st ACM Joint European Software Engineering Conference and Symposium on the Foundations of Software Engineering (ESEC/FSE)},
  pages = {1509--1521}
}

@inproceedings{zhang2024llmcommitstudy,
  author={Zhang, Linghao and Zhao, Jingshu and Wang, Chong and Liang, Peng},
  booktitle={Proceedings of 31st IEEE International Conference on Software Analysis, Evolution and Reengineering (SANER)}, 
  title={Using Large Language Models for Commit Message Generation: A Preliminary Study}, 
  year={2024},
  pages={126--130}
}

@inproceedings{xiong2025c3gen,
  author={Xiong, Bo and Zhang, Linghao and Wang, Chong and Liang, Peng},
  booktitle={Proceedings of the 19th ACM/IEEE International Symposium on Empirical Software Engineering and Measurement (ESEM)}, 
  title={Contextual Code Retrieval for Commit Message Generation: A Preliminary Study}, 
  year={2025},
  pages={358-364}
}

@article{xiong2025coracmg,
  title = {CoRaCMG: Contextual retrieval-augmented framework for commit message generation},
  journal = {Information and Software Technology},
  volume = {196},
  pages = {108169},
  year = {2026},
  author = {Bo Xiong and Linghao Zhang and Zongen Ren and Chong Wang and Peng Liang},
}

@article{xue2024llmcmg,
  author={Xue, Pengyu and Wu, Linhao and Yu, Zhongxing and Jin, Zhi and Yang, Zhen and Li, Xinyi and Yang, Zhenyu and Tan, Yue},
  journal={IEEE Transactions on Software Engineering}, 
  title={Automated Commit Message Generation With Large Language Models: An Empirical Study and Beyond}, 
  year={2024},
  volume={50},
  number={12},
  pages={3208--3224}
}

@inproceedings{kuang2025brevity,
  author = {Kuang, Hongyu and Zhang, Ning and Gao, Hui and Zhou, Xin and Assuncao, Wesley K. G. and Ma, Xiaoxing and Shao, Dong and Rong, Guoping and Zhang, He},
  title = {Brevity is the Soul of Wit: Condensing Code Changes to Improve Commit Message Generation},
  year = {2025},
  booktitle = {Proceedings of the 16th International Conference on Internetware (Internetware)},
  pages = {389--401}
}

@article{zhang2024ragenhancedcommitmessagegeneration,
      title={RAG-Enhanced Commit Message Generation}, 
      author={Linghao Zhang and Hongyi Zhang and Chong Wang and Peng Liang},
      year={2024},
      journal={arXiv preprint arXiv:2406.05514},
}

@inproceedings{shen2016automatic,
  author={Shen, Jinfeng and Sun, Xiaobing and Li, Bin and Yang, Hui and Hu, Jiajun},
  booktitle={Proceedings of the 40th IEEE Annual Computer Software and Applications Conference (COMPSAC)}, 
  title={On Automatic Summarization of What and Why Information in Source Code Changes}, 
  year={2016},
  pages={103--112}
}

@article{Li2024onlydiff,
  author = {Li, Jiawei and Farag\'{o}, David and Petrov, Christian and Ahmed, Iftekhar},
  title = {Only diff Is Not Enough: Generating Commit Messages Leveraging Reasoning and Action of Large Language Model},
  year = {2024},
  volume = {1},
  number = {FSE},
  journal = {Proceedings of the ACM on Software Engineering},
  pages = {1--22}
}

@inproceedings{tian2022goodcommit,
  author = {Tian, Yingchen and Zhang, Yuxia and Stol, Klaas-Jan and Jiang, Lin and Liu, Hui},
  title = {What makes a good commit message?},
  year = {2022},
  booktitle = {Proceedings of the 44th International Conference on Software Engineering (ICSE)},
  pages = {2389--2401},
}

@misc{israel1992determining,
  author       = {Israel, Glenn D.},
  title        = {Determining Sample Size},
  year         = {1992},
  number       = {PEOD-6},
  institution  = {Florida Cooperative Extension Service, Institute of Food and Agricultural Sciences, University of Florida}
}

@misc{replpack,
  author       = {Zongen Ren and Wei Shi and Bo Xiong and Chaoran Cai and Chong Wang and Peng Liang},
  title        = {{Replication Package for the Paper: LLM-Enhanced Commit Message Generation via Issue Information: An Exploratory Study}},
  year         = {2026},
  howpublished = {\url{https://github.com/Rnn123/ISAC-CMG}}
}

@misc{deepseek-api-docs,
  author       = {{DeepSeek}},
  title        = {{DeepSeek API Documentation}},
  year         = {2026},
  howpublished = {\url{https://api-docs.deepseek.com/}},
  note         = {Accessed: 2026-08-05}
}

@misc{openrouter-api-docs,
  author       = {{OpenRouter}},
  title        = {{OpenRouter API Documentation}},
  year         = {2026},
  howpublished = {\url{https://openrouter.ai/docs/api-reference/overview}},
  note         = {Accessed: 2026-08-05}
}

\end{document}